\documentclass[a4paper,11pt]{article}
\usepackage{jcappub}
\usepackage{multirow}

\arxivnumber{2407.05273}
\title{A Natural Explanation of the VPOS from Multistate Scalar Field Dark Matter}

\author[a,1]{Tula Bernal,\note{Corresponding author.}}
\affiliation[a]{\'Area de F\'isica, Depto.~de Preparatoria Agr\'icola, Universidad Aut\'onoma Chapingo, Km 38.5 Carretera M\'exico-Texcoco, Texcoco 56230, Edo. M\'ex., M\'exico}   

\author[b,2]{Tonatiuh Matos\note{Part of the Instituto
       Avanzado de Cosmolog\'{\i}a (IAC) Collaboration.}}
\affiliation[b]{Departamento de F\'{\i}sica, Centro de Investigaci\'on y de Estudios Avanzados del IPN, AP 14-740, Ciudad de M\'exico 07000, M\'exico}

\author[c]{and Leonardo San.-Hernandez}
\affiliation[c]{Departamento de F\'isica, Universidad Aut\'onoma Metropolitana Iztapalapa, San Rafael Atlixco 186, CP 09340,
Ciudad de M\'exico, M\'exico}

\emailAdd{tbernalm@chapingo.mx}

\abstract{Observations with the Gaia satellite have confirmed that the satellite galaxies of the Milky Way are not distributed as homogeneously as expected. The same occurs in galaxies such as Andromeda and Centaurus A, where satellites around their host galaxies have been observed to have orbits aligned perpendicular to the galactic plane of the host galaxy. This problem is known for the Milky Way as Vast Polar Structure (VPOS). The Scalar Field Dark Matter Field (SFDM), also known as Ultralight-, Fuzzy-, BEC-, and Axion-dark matter, proposes dark matter is a scalar field, which in the non-relativistic limit follows the Schr\"odinger equation coupled to the Poisson equation. Although the SF  here is classical, the Schr\"odinger equation contains a ground and excited states as part of its nature. In this work, we show that such quantum character of the SFDM can naturally explain the VPOS observed in galaxies. By taking into account the finite temperature corrections for a complex, self-interacting SF at very early epochs of the Universe, we show that with the ground and first excited states in the Newtonian limit, we can fit the rotation curves of the host galaxies. With the best-fit parameters obtained, we can explain the VPOS. We do this with particular galaxies, such as the Milky Way, Andromeda, Centaurus A, and 6 other galaxies whose satellites have been observed. This result shows that the multistate SFDM is not distributed homogeneously around the galaxy, and therefore might explain the anisotropic distribution of the satellite galaxies. According to this result, this could be a general characteristic of the galaxies in the Universe. Finally, we also show how the scale of each galaxy depends on a parameter determined by the final temperature of the SF galactic halo under study. This might explain why different galaxies with SFDM give different values of the mass of the SF.}

\begin{document}
\maketitle
\flushbottom

\section{Introduction}
\label{sec:introduction}

Dark Matter (DM) stands as one of the most important scientific puzzles of the current century. Despite advances, over 95\% of the Universe's matter-energy content eludes comprehension, with DM comprising more than a quarter of this total. DM is responsible for the formation of diverse structures in the Universe, ranging from superclusters of galaxies to dwarf galaxies scattered across the cosmos. Since its discovery in the early 1930s, unraveling the composition of dark matter has remained a significant challenge, representing one of the foremost enigmas in science.

The most widely accepted and studied model that explains various cosmological observations, as well as those on large galaxies and galaxy cluster scales, is cold dark matter (CDM). However, CDM presents some problems at galactic scales, for example the so-called ``core-cusp'' problem, the ``missing satellites'' problem, and all CDM models predict that satellite galaxies should move in homogeneous orbits within their host galaxies and be uniformly distributed \citep[see e.g.][]{Primack:2012id,DelPopolo:2016emo,Bullock:2017xww,Salucci:2020nlp,Peebles:2024txt}. However, observations from current telescopes have revealed the satellite galaxies of three galaxies, the Milky Way (MW) \citep{Pawlowski:2013cae,Pawlowski:2019bar}, Andromeda (M31) \citep{Ibata:2013rh,Conn:2013iu}, and Centaurus A (Cen A) \citep{Muller:2018hks,Muller:2020njt}, are not evenly distributed but exhibit a phenomenon known as Vast Polar Structure (VPOS), named in this way for the MW. Although the CDM model appears to be transiently compatible with this observation \citep{Sawala:2022xom}, it is challenging to explain this phenomenon in multiple galaxies, especially the coincidence that these three galaxies in our neighborhood exhibit this phenomenon.
Thus, the CDM paradigm faces challenges in its predictions, particularly on galactic scales \citep{Weinberg:2015}. This has motivated the search for alternative models consistent with observational data \citep[see e.g.][]{Oks:2021hef}.

One alternative is the Scalar Field Dark Matter (SFDM) model \citep[see e.g.][]{Matos:2000ss,Hui:2016ltb,Suarez:2013iw}, proposed first in \citep{Ji:1994xh,Lee:1995af}, and systematically studied the next years by many groups \citep[see e.g.][]{Matos:1998vk,Urena-Lopez:2002nup,Matos:2023usa}. The SFDM model proposes that DM is a scalar field (SF), a spin-0 particle, governed by the Klein-Gordon equations driving the dynamics of the Universe \citep[see e.g.][and references therein]{Matos:2023usa,Hui:2016ltb,Bernal:2017oih}. The SFDM model is known by different names, including Fuzzy \citep{Hu:2000ke}, Bose-Einstein \citep{Boehmer:2007um,Rindler-Daller:2009qyu}, Wave DM \citep{Bray:2010fc,Schive:2014dra}, among others. SFDM has gained popularity as one of the favored candidates to explain DM.

This hypothesis can explain the observed rotation curves of stars and gas around galaxies \citep[see e.g.][among others]{Matos:1998vk,Bernal:2017oih,Solis-Lopez:2021tou,Alvarez-Rios:2024}. Also, the SFDM model can explain all the cosmological observations as well as the CDM model; it reproduces the observed mass power spectrum and the angular power spectrum \citep[see e.g.][and references therein]{Matos:2000ss,Schive:2014dra,Hlozek:2014lca}. The difference with CDM is that, given its quantum nature, the SFDM model presents ``core'' density profiles instead ``cusp'' ones, and SFDM presents a natural cut-off in the mass power spectrum, implying that the theoretical number of satellite galaxies matches the observed quantities, solving in this way the ``missing satellites'' problem. 

Diverse proposals for the origin and nature of SFDM highlight the ongoing exploration and research efforts to comprehend better the mysterious nature of DM and its potential connections to fundamental physics. For instance, SFDM can arise from superstring theory \citep{Hui:2016ltb}; in this scenario, the SF is real and electrically neutral. Another possibility is to incorporate the SFDM Lagrangian into the Standard Model (SM) of particle physics, establishing a connection between DM and the known particle physics of the Universe. In \citep{Rindler-Daller:2013zxa}, a complex scalar field $\Phi$ is considered a potential candidate for SFDM. In \citep{Hernandez:2023tig}, the authors examine the possibility that SFDM consists of charged dark bosons whose gauge charge is associated with a dark photon.

The SFDM model has also faced some problems that have been solved. The first was discovered in \cite{Guzman:2003kt}, revealing the potential instability of galaxies under this model. The second issue concerned the possibility of having supermassive black holes at the center of galaxies consuming the entire SF halo. The latter problem was addressed showing that supermassive black holes can coexist with the SFDM halo \citep{Urena-Lopez:2002nup,Barranco:2011eyw,Avilez:2017jql,Padilla:2020sjy}.

The first problem was addressed by considering the quantum character of the SF taking into account multiple wave functions as states or excited states of the system. In reference \citep{Urena-Lopez:2010zva}, the properties of gravitationally bound multistate configurations composed of spin-zero bosons in the Newtonian regime were studied. They found that the system remains stable when containing multiple states simultaneously. This result was corroborated in \citep{Guzman:2019gqc} and more recently in \citep{Guzman:2022vxl}. This finding is particularly important for the present work. The multistate solutions are also known as $l$-boson stars \cite{Alcubierre:2018ahf}. 

The multistate SFDM has also been studied using quantum field theory at finite temperature in \citep{Matos:2011pd,Robles:2012kt,Robles:2013ioa}, for a real SF. This is the alternative that we choose to explore in the present work, but for a complex SF.
Our hypothesis involves a complex SF immersed in a thermal bath at finite temperature with the other SM particles, at the very early Universe. The SF is governed by a quartic SFDM potential, going through an spontaneous symmetry breaking (SSB) or phase transition, allowing the formation of SFDM halos with particles in a Bose-Einstein condensate ground state, and excited states also present, treating them as atom-like systems. We obtain an effective SF mass that depends on the temperature of the thermal bath and the critical temperature of the halo at the moment the system collapses.

We assume that a complex SF is the DM of the Universe and is the dominant component contributing to the gravitational field. The electromagnetic fields are assumed to be small and have a negligible contribution to the overall gravity of the system.
In a galaxy context, the SFDM can be effectively described by the non-relativistic limit of the Klein-Gordon (KG) equations. By applying this limit to the Einstein-KG field equations, they reduce to the Schrödinger-Poisson (SP) system, which describes the behavior of SFDM halos in a non-relativistic, quantum mechanical framework.

In this frame, we can interpret the SFDM halos of galaxies as macroscopic gravitational atoms with various quantum states similar to those observed in atomic physics, such as $s$, $p$, $d$, etc. The $s$ states correspond to spherical configurations, while the $p$ states resemble lobes or bubbles at the north and south poles of the galactic atom.
Numerical simulations and evolution of the SP system performed in \cite{Guzman:2019gqc} have shown that the SF density profile exhibits a resemblance to the Legendre functions $P_k^j(\cos\theta)$ of the hydrogen atom, where $k$ and $j$ are quantum numbers. This similarity indicates that the SFDM profile behaves similarly to a hydrogen atom, supporting the analogy between galaxies and atoms.

This quantum behavior of the SFDM model is the main difference between the SFDM paradigm and other DM models. This feature is just the main point that allows us to explain various DM observations, especially at galactic scales, in particular the anisotropic distribution of satellite galaxies observed in the Milky Way, known as Vast Polar Structure (VPOS) \citep{Pawlowski:2013cae,Pawlowski:2019bar}, Andromeda \citep{Ibata:2013rh,Conn:2013iu}, and Centaurus A \citep{Muller:2018hks,Muller:2020njt}, which conventional CDM models do not explain. 
Thus, taking into account the quantum behavior of the SFDM, we can explain the behavior of the VPOS in a very simple and natural way, as proposed first in \citep{Solis-Lopez:2019lvz,Thesis:Cristian}, which is the aim of this work \citep[see also][]{Jaramillo:2024smx}.

To do this, the article is organized as follows. In Section \ref{sec2:model}, we provide a summary of the finite temperature SFDM model, presenting in \ref{sec2a:previous} the previous work done for a real SF, the successes of such model, and unsolved issues. In subsection \ref{sec2b:ssb}, we start with the quartic potential for a complex SF, immersed in a thermal bath with the other components of the SM at very early stages of the Universe, and show the conditions of an SSB as the Universe cools down. In this subsection, we obtain the effective mass of the SF, which depends not only on the intrinsic mass of the field and its self-interaction parameter but also on the temperature of the halo at the moment the system gravitationally collapses and on the critical temperature at which the SSB occurs.
In \ref{sec2c:Newtonian}, we present the evolution equations of the complex SF after the SSB takes place in an FLRW Universe, separating the SF function into its background part and a linear perturbation, and show the Newtonian limit to find the analytic solutions for the density, as we are interested in already formed galactic halos.
In subsection \ref{sec2d:hydrodynamic}, we perform a polar decomposition of the SF function into its norm and its phase. This separation allows us to rewrite the KG equations as a continuity and Bernoulli equations. This is the hydrodynamic representation of the KG equations. This form of the KG equations allows us to point out that the only difference between a classical hydrodynamical system and the KG equation is a term known as the quantum potential. If such a quantum term is zero, the system is classical. Therefore, we can say that the quantum character of the KG equation is contained in this term. In subsection \ref{sec2e:solution} we find exact solutions for the ground and first excited states through the separation of variables, spherical harmonics, and Bessel polynomials, and we write the corresponding densities of those states. The main result shown here is that the final collapse of the SFDM is a halo in the form of an atom, with a first excited state in the form of two lobes along the north-south direction. To show if these resulting halos can explain the anisotropic distribution of the satellites in the MW, M31, CenA, and another 6 MW-like galaxies, we fit the exact mSFDM solutions to the observed rotation curves of the galaxies and used the best-fit parameters to compare the resulting atom-like structures with the gravitational potential needed to explain the distribution of the satellites around the galaxies. In Section \ref{sec3:galaxies}, we describe the bulge and disc models we use to fit the rotation curves of the galaxies. In Section~\ref{sec:mcmc}, we explain the MCMC calibration method we use to fit the models with the observations. In Section \ref{sec4:obs} we describe the observations we used, and the priors for the calibration method. In Section \ref{sec5:results} we show the results of the fittings and the resulting structures of the two-lobe halos. Finally, in Section \ref{sec:conclusions} we present some conclusions.

\section{Finite Temperature SFDM Model}
\label{sec2:model}

At the beginning of the Universe, we assume that SFDM was in thermal equilibrium with the other components of the Standard Model (SM) of particles but decoupled from them very early in the Universe's history. Starting with the SF quartic potential $V$ coupled to a thermal bath with temperature $T$, we have \citep{Kolb-Turner:1990, Dalfovo:1999}
\begin{equation}
\label{eq:V}
    V =-m^2\Phi\Phi^*+\frac{\lambda}{2}(\Phi\Phi^*)^2+\frac{\lambda}{4}\Phi\Phi^*T^2+\frac{\pi^2}{90} T ^4 ,
\end{equation}
for the case when $T \gg m$, where $m$ is the mass of the SF and $\lambda$ the self-interaction parameter between SF particles. Observations indicate that the self-interaction strength $\lambda$ must be extremely small but non-zero, as constrained by nucleosynthesis \citep{Li:2013nal}. The mass $m$ of the SFDM is considered ultralight, which implies that any perturbation in its temperature is comparable to this mass in energy.

Now, of particular interest for cosmology is the theoretical expectation that at high temperatures, symmetries that are spontaneously broken today were restored and, during the evolution of the Universe, there were phase transitions associated with the spontaneous breakdown of gauge symmetries \citep{Kolb-Turner:1990}. So we expect that, as the Universe expanded, the SFDM particles and the other SM components, cooled down and underwent a phase transition at a critical temperature $T_c$ very early in the cosmic time. At this point, a spontaneous symmetry breaking (SSB) of the SF potential takes place.

According to the standard cosmological model, after inflation, quantum fluctuations evolved into classical fluctuations, leading to the collapse of the SFDM and the formation of cosmic structures \citep{Matos:2000ss, Urena-Lopez:2010zva,Hlozek:2014lca}. As the Universe expanded, the volume of these fluctuations increased, causing the SF to cool and form Bose-Einstein condensates. These condensates subsequently formed galaxy halos, where most SFDM particles settled into the ground state. However, gravitational forces caused the fluctuations to collapse again, decreasing their volume and increasing the halo's temperature, resulting in some SFDM particles transitioning to excited states.

In reference \cite{Guzman:2019gqc}, it was demonstrated that if the majority of SFDM particles remain in the ground state while a portion of them occupy at least one excited state, the system stabilizes and remains stable for at least the age of the Universe, if the mass of the SFDM particle is sufficiently small, about $m \approx 10^{-22}\text{eV}/c^2$. This stability mechanism, involving a mixture of ground and excited states for SFDM particles within halos, is crucial for ensuring the formation and persistence of cosmic structures.

\subsection{Previous Work and Unsolved Issues}
\label{sec2a:previous}

\citep{Robles:2012kt} studied the case of a real SF with mass $m$ and self-interaction $\lambda$, in a thermal bath with temperature $T$. Starting from the SF potential \eqref{eq:V}, it was shown that the characteristic polynomial for this system is given by
\begin{equation}
    \omega^2 = k^2 c^2 + m^2 c^2 \left(1 - \frac{T^2}{T_c^2}\right) ,
\end{equation}
where $\omega$ corresponds to the angular frequency of the temporal part of the solution, $k$ to the wave number of the radial part, $c$ is the speed of light and $T_c=2m/\sqrt{\lambda}$ is its critical phase-transition temperature.

As the SF in a galaxy halo heats up, the effective mass $m_\Phi$ decreases, following the relation
\begin{equation}
    m_\Phi^2 = m^2 \left(1 - \frac{T^2}{T_c^2} \right) .
\label{eq:m-realSF}
\end{equation}
Consequently, galaxies of different sizes exhibit different effective masses $m_\Phi$. For example, if we assume that the mass of the scalar field is $m\sim 10^{-21} \text{ eV}/c^2$, which agrees with the constrictions of Lyman alpha observations \citep{LinaresCedeno:2020dte} and of the satellite galaxies of the Milky Way \citep{DES:2020fxi}, the effective SF mass is of the order of $m_\Phi \sim10^{-24} \text{ eV}/c^2$. The halo of such a galaxy remains as a Bose-Einstein condensate, preserving its DM behavior. However, with the increase in temperature, some SFDM particles should occupy excited states, and the effective mass obtained from different observations should be different, solving a current problem found with the Fuzzy DM model \citep[see e.g.][]{Ferreira:2020fam,Bernal:2017oih}.

In \citep{Robles:2012kt}, the finite temperature density profile is obtained as
\begin{equation}
    \rho(r) = \sum_j \rho_0^j \frac{\sin^2 (j \pi r/ R)}{(j \pi r/R)^2} ,
\label{eq:rho-realSF}
\end{equation}
with $\rho_0^j$ the central density of the multistate $j$ and $R$ the radius of the whole SFDM configuration. In this case, the multistate SFDM halo is the sum of the same multistates' functional form with different phases.

The last multistate solution was shown capable of fitting the rotation curves of low and high surface brightness galaxies and the wiggles observed in the rotation curves of some of them \citep{Bernal:2017oih}, the strong gravitational lensing in SFDM halos \citep{Robles:2013dfa}, reproduced successfully the evolution of a dwarf satellite galaxy embedded in the finite temperature SFDM halo in numerical simulations \citep{Robles:2014ysa} and reproduced the dynamical mass distribution in clusters of galaxies \citep{Bernal:2016lll}.

A disadvantage of this model is that different combinations of ground plus excited states are required to reproduce the observations, which could be explained due to the distinct formation processes of galaxies, groups, and clusters of galaxies, resulting in different numbers of SF particles in diverse states.


\subsection{Spontaneous Symmetry Breaking with a Complex Scalar Field}
\label{sec2b:ssb}

For the present article, we start with the following SF potential $V$ at the early stages of the Universe, for a complex SF, $\Phi$:
\begin{equation}
    V=\frac{\lambda}{2}\left(\Phi\Phi^*-\frac{m^2}{\lambda}+\frac{1}{2}T^2\right)^2 ,
\label{eq:V2}
\end{equation}
which is the same as potential \eqref{eq:V} plus some gauge terms that depend on the SF mass $m$ and the temperature of the thermal bath $T$.
We will denote $\Phi\Phi^*:=n$ as the SF number density in what follows. From the first derivative of equation~\eqref{eq:V2}, we find that this potential has the following two extreme points:
\begin{equation}
\begin{split}
    \Phi_1 \Phi_1^* &= n_1 = 0 ; \\
    \Phi_2 \Phi_2^* &= n_2 =\frac{m^2}{\lambda}-\frac{1}{2}T^2 .
\end{split}
\label{eq:points}
\end{equation}
 
\begin{figure}
    \centering
    \includegraphics[scale=0.55]{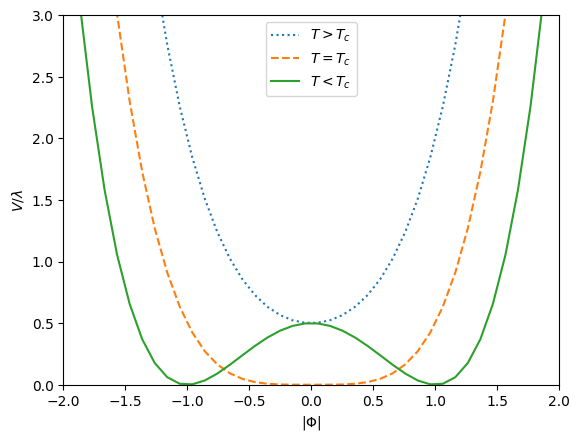}
    \caption{Qualitative behavior of the SF potential~\eqref{eq:V2} for three values of the temperature $T$ with respect to the critical temperature $T_c$. The dotted blue line corresponds to $T > T_c$, where the system oscillates around the minimum of the potential. The dashed yellow line corresponds to $T=T_c$ where the system undergoes the SSB. The solid green line $T<T_c$ corresponds to the new potential with two minima and one maximum point}
    \label{fig:ssb-potential}
\end{figure}

The second derivative reads 
\begin{equation}
    V_{,\Phi\Phi^*}=\lambda\left(2n-\frac{m^2}{\lambda}+\frac{1}{2}T^2\right) ,
\label{eq:V2nd}
\end{equation}
which takes the value 
$V_{,\Phi\Phi^*}|_{n_1}=-m^2+ \lambda T^2/2$ for the first critical point in \eqref{eq:points}, and $V_{,\Phi\Phi^*}|_{n_2}=m^2- \lambda T^2/2$ for the second one. If we define:
\begin{equation}
    T^2_c := \frac{2m^2}{\lambda} ,
\end{equation}
the values of the second derivative \eqref{eq:V2nd} read $V_{,\Phi\Phi^*}=\pm(\lambda/2) \left( - T_c^2 + T^2 \right)$.
This implies that the first extreme point is a minimum for $T^2>T_c^2$, and it is a maximum for $T^2<T_c^2$. The second extreme point does not exist for $T^2>T_c^2$ (since the corresponding $n_2<0$), and it is a double minimum for $T^2<T_c^2$ (see figure~\ref{fig:ssb-potential}). That means that $T_c$ is the critical temperature where the SSB takes place and the SF potential~\eqref{eq:V2} changes from one minimum point to two minima and one maximum point.
For the first critical value, $n_1=0$, the SF potential takes the value
$V(n_1)=(\lambda/2) \left(-\frac{m^2}{\lambda}+\frac{T^2}{2}\right)^2$, and for the second one, $n_2$, the SF potential is $V(n_2)=0$. Observe that $n=0$ at the critical temperature $T_c$.

Now, the effective mass $m_\Phi$ of the SF is defined as
$m_\Phi^2 := \left. V_{,\Phi\Phi^*} \right|_{\text{min}}$, so we obtain:
\begin{eqnarray}
\label{eq:Mass}
    m_\Phi^2 = \left\{\begin{matrix}
    -m^2+\frac{\lambda}{2}T^2=\frac{\lambda}{2}(T^2-T_c^2) & \text{for}\,\,\, n_1; \quad T^2>T_c^2 ,\\ \\
     m^2-\frac{\lambda}{2}T^2=\frac{\lambda}{2}(T_c^2-T^2) & \text{for}\,\,\, n_2 ; \quad T^2<T_c^2 ,
    \end{matrix}
    \right.
\end{eqnarray}
therefore $m_\Phi^2$ is always positive. 
As in the case of the real SF (equation~\eqref{eq:m-realSF}), notice from the last result that the effective mass of the complex SF depends on the temperature $T$ of the SF at the moment the halo collapsed, and the critical temperature $T_c$, which depends itself on the mass $m$ and the self-interaction parameter $\lambda$, intrinsic properties of the SF particle. This implies that the effective observed mass will depend on the particular properties of the observed system. Again, this may explain the tensions on the SF mass obtained from the fitting with diverse astrophysical and cosmological observations for the fuzzy DM model \citep[see bellow and e.g.][]{Ferreira:2020fam,Bernal:2017oih}.

\subsection{Evolution Equations and Newtonian Limit}
\label{sec2c:Newtonian}

Now, we study the evolution of the SF with the potential \eqref{eq:V2} and finite temperature, in a FLRW Universe.
The SF fulfills the Einstein-Klein-Gordon equations (EKG), $G_{\mu\nu}=\kappa^2T_{\mu\nu}$, and $\Box\Phi-\frac{dV}{d\Phi^*}=0$, where $G_{\mu\nu}$ is the Einstein tensor, $T_{\mu\nu}$ is the energy-momentum tensor of the SF and $\kappa^2:=8\pi G/c^4$. If we separate, as usual, the background part from a perturbed fluctuation, $\Phi=\Phi_0+\delta\Phi$, the fluctuation $\delta\Phi$ fulfills the following equations:
\begin{equation}
    \nabla^2\delta\Phi-\ddot{\delta\Phi}-2H\dot{\delta\Phi} +V_{,\Phi_0\Phi_0^{*}}a^2\delta\Phi
    +V,_{\Phi^{*}_0\Phi^{*}_0}a^2\delta\Phi^{*}-2V,_{\Phi^{*}_0}a^2\phi+4\dot{\phi}\dot\Phi_0 = 0,
\label{eq:KGP}
\end{equation}
\begin{equation}
    2\nabla^2\phi-6H(\dot{\phi}+H\phi) =
    \kappa^2\left[(\dot{\Phi}_0\dot{\delta\Phi}^{\ast}+\dot{\Phi}^{\ast}_0\dot{\delta\Phi})-2\phi\dot{\Phi}_0\dot{\Phi}^{\ast}_0+a^2\delta V \right],
\label{eq:Poisson}
\end{equation}
where $H$ is the Hubble parameter, $a$ the scale factor, $\phi$ the Newtonian potential, and $\delta V$ is defined as
\begin{equation}\label{eq:deltaV}
    \delta V := V,_{\Phi_0} \delta\Phi + V,_{\Phi_0^{\ast}}\delta\Phi^{\ast}.
\end{equation}
Equation~\eqref{eq:KGP} is the KG equation for the perturbation $\delta \Phi$, and equation \eqref{eq:Poisson} is the Poisson equation.

Now, we aim to study the final formation stage of galactic halos, thus, we constrain our evolution equations to the Newtonian limit, when $\Phi$ is near the minimum of the potential and it starts to oscillate. In this limit, we can neglect the expansion of the Universe, i.e. $H=0$, and the oscillations of the Newtonian gravitational potential are also neglected, i.e. $\dot\phi=0$, since the gravitational potential is locally homogeneous at the beginning of the collapse. 

For the SF potential \eqref{eq:V2}, equations \eqref{eq:KGP} and \eqref{eq:Poisson} read
\begin{equation}
    \nabla^2\delta\Phi-\ddot{\delta\Phi}+\lambda\left(2n_0-\frac{m^2}{\lambda}+\frac{1}{2}T^2\right) a^2\delta\Phi +\lambda\Phi_0^2 a^2\delta\Phi^{*} = 0,
\label{eq:KGP1}
\end{equation}
\begin{equation}
    \nabla^2\phi =
    \frac{\kappa^2}{2} \left[ \left( \dot{\Phi}_0\dot{\delta\Phi}^{\ast}+\dot{\Phi}^{\ast}_0\dot{\delta\Phi} \right) - 2\phi\dot{\Phi}_0\dot{\Phi}^{\ast}_0 + a^2\lambda\left(\Phi_0\Phi_0^*+\frac{m^2}{\lambda}-\frac{1}{2}T^2\right) \left( \Phi_0\delta\Phi^*+
\Phi_0^*\delta\Phi \right) \right] .
\label{eq:Poisson1}
\end{equation}

\subsection{Hydrodynamic Representation}
\label{sec2d:hydrodynamic}

To solve the last equations, we carry out a Madelung transformation \cite{Matos:2022quantum}:
\begin{eqnarray}
\label{eq:deltaPhi2}
    \Phi_0&=&\sqrt{n_0}\,e^{i\theta_0}=\sqrt{n_0}\,e^{i(S_0-\omega_0 t)},\\
    \delta\Phi&=&R\,e^{i(\theta_0+\delta S)}=\sqrt{n}\,e^{i(\theta_0-\omega_0 t+\delta S )},
\label{eq:deltaPhi3}
\end{eqnarray}
where $\Phi$ is decomposed into a density number of the SF $n$ and a phase $S$.
We will take perturbations such that if $\phi$ is of the order of $\epsilon$, then $\delta\Phi$, $\delta R$, and $\delta S$ are also of this order, while $n$ is a perturbation of order $\epsilon^2$. We will neglect perturbations of order $\epsilon^2$ and beyond. With these new variables, equations \eqref{eq:KGP1} and \eqref{eq:Poisson1} transform into
\begin{equation}
\label{eq:KGPn0}
    \nabla^2 R-\ddot{R}
    +\frac{\lambda}{2}\left(4n_0-T^2_c+T^2\right)a^2R+ \dot\theta_0^2R-i(\dot\theta_0R^2\dot)=0,
\end{equation}
\begin{equation}
\label{eq:Poissonn0}
    \nabla^2\phi=
    \frac{\kappa^2}{2}\left[-\left(2\dot\theta_0^2n_0+\frac{\dot n_0^2}{2n_0}\right)\phi +\frac{\dot n_0}{\sqrt{n_0}}\dot R+\left(\lambda(2n_0+T_c^2-T^2)a^2+2\dot\theta_0^2\right)\sqrt{n_0}R\right].
\end{equation}
After the formation of the galaxy halo, the size of the final configuration determines the final temperature $T$.

\subsection{Exact Solution}
\label{sec2e:solution}

To solve the last field equations, we perform a separation of variables defining $R:=L_kY_k^jT_R$, where $L_k=L_k(r)$, $T_R=T_R(\eta)$ and $Y_k^j=Y_k^j(\theta,\varphi)$ are the spherical harmonics polynomials, with $r$ the radial, $\eta$ the temporal, and $\theta$ and $\varphi$ the angular coordinates. The real part of equation (\ref{eq:KGPn0}) has the solution
\begin{equation}\label{eq:R}
    R=\sum_{k,j}\frac{J_{k+1/2}(l\,r)}{\sqrt{r}}Y^j_k T_R ,
\end{equation}
where $J_k=L_k\sqrt{r}$ are the Bessel polynomials, and the function $T_R$ is a solution of the equation
\begin{equation}
\label{eq:TR}
    \ddot T_R + \left( \omega^2-2\lambda n_0a^2-\dot\theta_0^2 \right)T_R = 0,
\end{equation}
with $l$ an integration constant such that
\begin{equation}
\label{eq:l}
    l^2=\omega^2-\frac{\lambda}{2}\left(T^2_c-T^2\right)a^2=\omega^2-m_\Phi^2a^2 ,
\end{equation}
where we used the definition of the SF mass \eqref{eq:Mass}. The last equation is crucial for what follows. This equation means that $l$ is a parameter that depends on the value of the effective mass $m_\Phi$ and the vibration energy $\omega$. We will see that $l$ determines the scale of the galaxy, i.e.~the scale $l$ depends strongly on the galaxy we are studying and this scale depends on the temperature and energy with which the galaxy ends after its formation collapse. In other words, the final state of the galaxy determines the temperature of the SFDM, and the scale of the galaxy is determined by the effective mass given by this temperature. By fitting the exact solutions to the rotation curves of galaxies, we can obtain a value for the scale $l$, but unfortunately, we cannot determine the final temperature of the SFDM halo, we can only have the combination $\omega^2-m_\Phi^2 a^2$.
Thus, we can say that equation (\ref{eq:l}) can solve the mass tension between the mass that fits Lyman-$\alpha$ observations and the mass needed to fit rotation curves in galaxies. Let us explain this point a bit further. It has been argued that the mass of the SF $m_\Phi$ appears to be different to fit observations of large galaxies, small galaxies, and observations of Lyman-$\alpha$ \citep{Hlozek:2014lca,Chen:2016unw,Hlozek:2017zzf,Safarzadeh:2019sre,Broadhurst:2019fsl,Rogers:2020ltq,Pozo:2020fft,Pozo:2020ukk}. It is clear from equation \eqref{eq:Mass} that after the turn-around of the galactic halo occurs, the SFDM changes its temperature because it collapses again. Depending on the fluctuation size, environmental conditions, etc., a final SFDM halo temperature will be reached, and hence an effective frequency $\omega$ and the effective SF mass $m_\Phi$. There are two possibilities here. The first is that the self-interactions parameter is of order $\lambda\sim 1$, and then if $T\sim T_c$, the SFDM mass $m_\Phi$ could be the one observed using Lyman-$\alpha$ observations, $m_\Phi \sim 10^{-21}$eV, and the effective mass of a medium-sized galaxy given by the equation \eqref{eq:Mass} could be $m_\Phi\sim10^{-22}$eV.
The second possibility is that the self-interaction parameter $\lambda \ll 1$, as claimed in \citep{Li:2013nal}. In that case, $T^2/T_c^2\sim 0$ and it is sufficient that $m_\Phi\sim \omega$ for the scale $l$ to fit well to any galaxy. In other words,
equation \eqref{eq:l} can relieve the mass tension of the SFDM model, and the scale $l$ of a galaxy is determined not only by the mass $m_\Phi$ but also by the parameter $\omega$.

For equation \eqref{eq:Poissonn0}, we can expand the Newtonian potential as $\phi=pT_pY^j_k$, where again $p=p(r)$, $T_p=T_p(\eta)$ and $Y_k^j=Y_k^j(\theta,\varphi)$. Such an equation transforms into
\begin{equation}\label{eq:p}
    \frac{1}{r^2}\left(r^2 p_{,r}\right)_{,r}+\left(\Omega_1^2-\frac{k(k+1)}{r^2}\right)p=\Omega_2^2 L_k ,
\end{equation}
whose solution is
\begin{equation}
    p=\sum_{k,j}\left(\frac{J_{k+1/2}(\Omega_1\,r)}{\sqrt{r}}+\Omega_2^2 p_0\frac{J_{k+1/2}(l\,r)}{\sqrt{r}(\Omega_1^2-l^2)}\right)
    \label{eq:Solp}
\end{equation}
where $p_0$ is an integration constant. It can be seen that solution (\ref{eq:Solp}) is well behaved even at $\Omega_1=l$. Functions $\Omega_i=\Omega_i(\eta)$ are defined as
\begin{equation}
    \Omega_1^2 = \frac{\kappa^2}{2}\left(2\dot\theta_0^2n_0+\frac{\dot n_0^2}{2n_0}\right)=\kappa^2\dot\Phi_0\dot\Phi_0^*,
\end{equation}
\begin{equation}
    \Omega_2^2 = \frac{\kappa^2}{2}\left[\frac{\dot n_0}{\sqrt{n_0}} \frac{\dot{T}_R}{T_p}+\left(\lambda(2n_0+T_c^2-T^2)a^2+2\dot\theta_0^2\right)\sqrt{n_0}\frac{T_R}{T_p}\right].
\end{equation}
One interesting solution is when 
\begin{equation}
    T_p = \frac{\kappa^2}{2}\left[\frac{\dot n_0}{\sqrt{n_0}} {\dot{T}_R}+\left(\lambda(2n_0+T_c^2-T^2)a^2+2\dot\theta_0^2\right)\sqrt{n_0}{T_R}\right],
\end{equation}
in that case $\Omega_2=1$.
Notice that the time dependence of the equations is due to the expansion of the Universe, which is too slow compared to the movements of the galaxies. Therefore, if we assume that such functions of the background are almost constant in time, as a good approximation we can set all these functions as constant. Thus, we can consider $\Omega_1$ and $\Omega_2$ as constants. 
In that case, equation \eqref{eq:p} can be solved analytically. Furthermore, equation \eqref{eq:TR} for the time evolution of function $R$ is $T_R=T_0\exp(i(\omega^2-2\lambda n_0a^2-\dot\theta_0^2)\eta)$, and the density number $n$ reduce to
\begin{equation}
    n=\sum_{k,j}\frac{J^2_{k+1/2}(l\,r)}{r}Y^j_kY^{*j}_k ,
\end{equation}
where we have used the orthogonal properties of the spherical harmonic functions.
 
From equation \eqref{eq:R}, using the orthogonality of functions $Y^j_k$, we can define the complex function $R_k$ as
 \begin{equation}
     R_k=\frac{J_{k+1/2}(l\,r)}{\sqrt{r}} \exp(i\omega_k\eta) .
 \end{equation}
For the ground state ($k=0$) and the first excited state ($k=1$), we have the following exact solutions:
\begin{equation}
  R_0 = \frac{J_{1/2}(lr)}{\sqrt{r}} \exp(i\omega_0\eta) = R_{00}\frac{\sin x}{x}\exp(i\omega_0\eta) ,
\label{eq:R0}
\end{equation}
\begin{equation}
   R_1 = \frac{J_{3/2}(lr)}{\sqrt{r}} \exp(i\omega_1\eta) = R_{10}\left(\frac{\sin x}{x^2}-\frac{\cos x}{x}\right)\exp(i\omega_1\eta) ,
\label{eq:R1}
\end{equation}
where we have defined $x:=lr$, and $R_{00}$ and $R_{10}$ are constants. 

\begin{figure}
    \centering
    \includegraphics[scale=0.3]{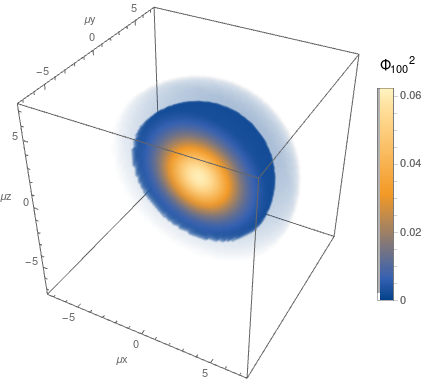}
    \includegraphics[scale=0.3]{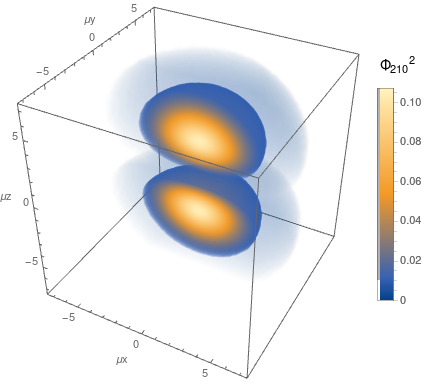}
    \includegraphics[scale=0.3]{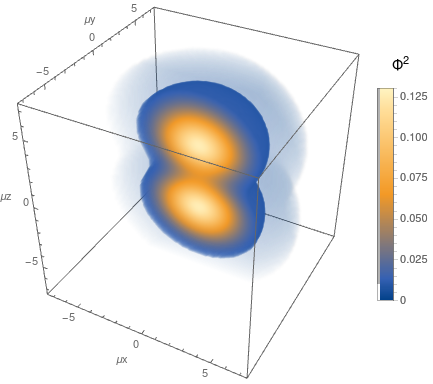}    
    \caption{(Left) A numerical simulation of the ground state of the SFDM. (Center) The same simulation but showing the evolution of the two-lobe excited state. (Right) The two states combined in the same image.}
    \label{fig:3D-states}
\end{figure}

See figure~\ref{fig:3D-states} for a 3D representation of the spherical ground state, the first excited state, and the sum of these two states of the SFDM.
The main result of this work is that we obtained, in an analytical way, a spherical ground state, as expected, and a first excited state with a two-lobe shape towards the north and south of the center of the configuration. This structure is like that of the hydrogen atom, called a gravitational atom. The two-lobe structure can explain why satellite galaxies in MW-like galaxies are not distributed spherically, but rather along a north-south line, which has been called VPOS. This behavior has not only been observed in the Milky Way \citep{Pawlowski:2013cae,Pawlowski:2019bar}, but also in Andromeda \citep{Ibata:2013rh,Conn:2013iu} and Centaurus A \citep{Muller:2018hks,Muller:2020njt}. In this way, our result has the potential to explain this anisotropic distribution of satellite galaxies. Additionally, we tested the model with the distribution of satellites in 6 other galaxies reported in \citep{Nashimoto_2022}.

This behavior, with a significant number of particles in excited states, while still maintaining the Bose-Einstein condensate nature of the halo, is essential for understanding the implications of the multistate SFDM model for galaxy formation and the large-scale structure of the Universe. It offers a mechanism for accommodating the observed stability of galaxies and their structures while accounting for the deviations in the effective mass due to temperature effects.

\begin{figure}
    \centering
    \includegraphics[scale=0.65]{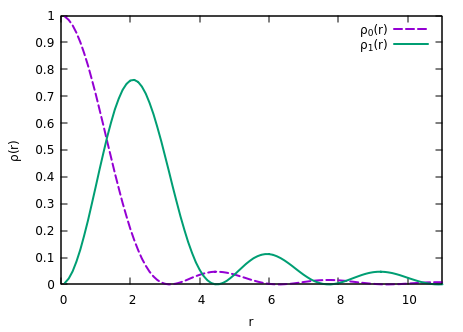}
    \caption{The purple dashed line is the density of the ground state, centered at $r=0$. The blue solid line is the density of the first excited state, with its first maximum outside $r=0$. Note that the ground state might decay very quickly, and the first excited state decays slower, even if its first maximum is smaller than the first maximum of the ground state.}
    \label{fig:2-states}
\end{figure}

Now, from \eqref{eq:R0} and \eqref{eq:R1}, the physical densities are given by $\rho_k (r) = R_k(r) R_k^*(r)$, and we define $\rho_0:=R_{00}^2$ and $\rho_1:=R_{10}^2$, with the constants $\rho_0$, the central density of the ground state, and $\rho_1$, the corresponding maximum density of the first excited state (see figure~\ref{fig:2-states}):
\begin{equation}
    \rho_0(r) = \rho_0 \frac{\sin^2 x}{x^2} ,
\label{eq:rho0}
\end{equation}
\begin{equation}
    \rho_1(r) = \rho_1 \left(\frac{\sin x}{x^2}-\frac{\cos x}{x}\right)^2 .
\label{eq:rho1}
\end{equation}
Notice that the density of the ground state \eqref{eq:rho0} is of the same form as in the case of the real SF solution (equation~\eqref{eq:rho-realSF}), for $j=1$ and $l=\pi/R$. In our case, the value $R=\pi/l$ defines the radius of the first minima of the ground state (see figure~\ref{fig:2-states}). In \citep{Robles:2012kt}, the authors defined $R$ as the radius of the SFDM halo. However, in this work, since there is still contribution to the mass beyond the first minimum, as seen in figure \ref{fig:2-states}, we define the maximum radius of every SF state, $R_i^\text{max}$, such that
\begin{equation}
    \rho_i(R_i^\text{max})=200\rho_c ,
\label{R-max}
\end{equation}
for $i=0,1$, where $\rho_c$ is the critical density of the Universe.

Another difference with the previous work \citep{Robles:2012kt}, is that the SFDM states in the case of the real SF are all centered at r=0, only their wavelength decreases and their amplitude changes, being all spherically symmetric. In that case, all the excited states share the definition of the radius $R$ of the halo. In the present work, for the complex SF, the second excited state has a different functional form, which gives it the two-lobe shape, with $\rho_1(0)=0$ and its maximum density outside $r=0$ (see figure~\ref{fig:2-states}).

The corresponding masses of both states, $M_i(r) = 4\pi\int \rho_i(r) r^2 \text{d}r$, are given by :
\begin{equation}
    M_0 (r) = \frac{2\pi
\rho_0}{l^3}\left(x-\cos x \sin x \right),
\end{equation}
\begin{equation}
    M_1 (r) = \frac{2\pi \rho_1}{l^3}\left(\frac{\cos(2x)}{x}-\frac{1}{x}+\cos x \sin x+x\right) .
\end{equation}

From now on, we will denote the resulting density profiles as multistate SFDM (mSFDM).
For the densities and masses, we have 3 free parameters only, $\rho_0$, $\rho_1$ and $l$. In the next section, we explain how we fit these functions to the rotation curves of the galaxies we are interested in, to obtain the halo configurations.

\section{Rotation Curves of Galaxies}
\label{sec3:galaxies}

To test this model and obtain the best values of the parameters $\rho_0$, $\rho_1$, and $l$ for the ground and the first excited states of the SFDM halo, respectively, we fit the rotation curves of the Milky Way, Andromeda, Centaurus A, as the main galaxies where the anisotropy of the satellite galaxies has been observed. Also, we fit the rotation curves of 6 MW-like galaxies whose satellite galaxies have been observed \citep{Nashimoto_2022}.

The rotation curve (the tangential velocity of the satellites) is given by $v^2(r)=GM(r)/r$, for $M$ the mass of the galaxy. In what follows, we fit the rotation curves of different galaxies with this formula, such that
\begin{equation}
    v(r) = \sqrt{ \frac{G}{r} \left( M_b + M_d + M_0 + M_1 \right) } ,
\label{eq:vel}
\end{equation}
where the subscripts are for the bulge, disc, ground, and first excited states, respectively. We do not consider the gas contribution.

To test the plausibility of the model in explaining the anisotropic distribution of satellite galaxies around a host galaxy, we fit the best models that reproduce the observed rotation curve of the main galaxy, with the ground state (spherically symmetric) and the first excited state (with two lobes) of the mSFDM as the DM halo.
As said before, we define the radius of every state such that $\rho_i(R_i)=200\rho_c$, for $i=0,1$, where $\rho_c = 1.283 \times 10^{-7} M_\odot \text{pc}^{-3}$ is the critical density of the Universe (equation \eqref{R-max}). Both states present density oscillations until reaching the density of $200\rho_c$. These halos are like concentric shells that contribute less and less to the gravitational potential.

We assume that the motion of stars and gas in the galaxies is circular, without considering non-axisymmetric structures such as the bar and spiral arms. For the Milky Way, we fit an exponential bulge, an exponential disc, and the mSFDM halo, through 7 free parameters. For Andromeda, Centaurus A, and NGC 3437, we fit an exponential disc and the mSFDM halo, avoiding the central regions, through 5 free parameters, with initial parameters close to those reported in the literature.

For the last 5 galaxies from the work of \cite{Nashimoto_2022} (NGC 2950, 3245, 3338, 5866, and 7332), we fit the corresponding rotation curves with mSFDM only. In these cases, we have 3 free parameters only, $\rho_0$, $\rho_1$, and $l$.

To fit the free parameters as explained before, we implemented a Markov Chain Monte Carlo (MCMC) code with the \texttt{emcee} library in Python \cite{emcee:2013}, as explained in Section~\ref{sec:mcmc}. We report the best-fit parameters and errors in Section~\ref{sec5:results}.

\subsection{Exponential Bulge Model}

For the bulge in the MW, we use the exponential-sphere model. In this case, the volume mass density $\rho_b$ is an exponential function of the radius $r$ with a scale radius $h_b$, as follows \citep{Sofue:2017}:
\begin{equation}
    \rho_b(r) = \rho_{b,0} e^{-r/h_b} ,
\end{equation}
with $\rho_{b,0}$ the central bulge density. The corresponding mass enclosed on a sphere of radius $r$ is
\begin{equation}
    M_b(r) = m_b \left[ 1 - e^{-y} \left( 1 + y + \frac{y^2}{2} \right) \right] ,
\end{equation}
where $y:=r/h_b$, while the total mass of the bulge, $m_b$, is
\begin{equation}
    m_b = \int_0^\infty 4 \pi r^2 \rho_b(r) \text{d} r = 8 \pi h_b^3 \rho_{b,0} .
\end{equation}
The bulge rotation curve is
\begin{equation}
    v_b(r) = \sqrt{\frac{GM_b(r)}{r}}.
\end{equation}

\subsection{Exponential Disc Model}
\label{disc-model}

For the galactic discs, we use the exponential Freeman model for the surface mass density \citep{Freeman:1970}:
\begin{equation}
    \Sigma_d(r) = \Sigma_{d,0} e^{-r/h_d} ,
\end{equation}
where $\Sigma_{d,0}$ is the central density and $h_d$ is the scale radius. The total mass is given by $m_d = 2 \pi \Sigma_{d,0} h_d^2$. The corresponding rotation curve is given by
\begin{equation}
   v_d(r) = \sqrt{4 \pi G y^2 \Sigma_{d,0} h_d \left[ I_0(y) K_0(y) - I_1(y) K_1(y) \right]} ,
\end{equation}
where $y:=r/(2h_d)$, and $I_i$ and $K_i$ (for $i=0,1$) are the modified Bessel functions.

\section{MCMC Calibration Method}
\label{sec:mcmc}

We use a Markov Chain Monte Carlo (MCMC) code to constrain the free parameters of the mSFDM model and find the best fit for the velocity rotation curve in each galaxy.
We implemented the affine invariant ensemble sampler \cite{affinMCMC} for MCMC using the \texttt{emcee} library in Python \cite{emcee:2013}, maximizing the likelihood function $\mathcal{L}(\mathbf{p})$ given by
\begin{equation}
	\mathcal{L}({\bf p}) = \frac{1}{(2 \pi)^{N/2}
	|{\bf C}|^{1/2}} \exp{\left ( - \frac{{\bf \Delta}^{T}
	{\bf C}^{-1} {\bf \Delta}}{2} \right )} ,
\label{eq:likelihood}
\end{equation}
where $\mathbf{p}$ is the vector of parameters, $N$ the number of
data points for each galaxy, ${\bf \Delta} = v_\mathrm{obs}(r_i) -
v(r_i,\mathbf{p})$, for $v_\mathrm{obs}$ the observational
circular velocity at radius $r_i$, and $v(r_i,\mathbf{p})$ the total velocity given by \eqref{eq:vel}, computed in the same position where $v_\mathrm{obs}$ was
measured, and $\mathbf{C}$ a diagonal matrix.

We sample the parameter space from uniform prior ranges using $50$ random walkers, each with $5 \times10^{4}$ steps. We tested the convergence of the fit by monitoring the mean integrated autocorrelation time $\tau_\text{int}$ at each step. Convergence is considered achieved when the number of steps $N_\text{step}$ is at least 100 times $\tau_\text{int}$, and $\tau_\text{int}$ reaches an approximately steady state, fluctuating by less than $1\%$ per step (see references \cite{Hogg_2018, affinMCMC, emcee:2013}, for a complete discussion of the use of autocorrelation time as a convergence criterion).

We estimate the best-fit parameters by calculating the median of the posterior distribution associated with each parameter. We present the $1D$ histograms and $2D$ contours of the posterior distributions, as well as the value $-2 \ln \mathcal{L}$ associated with the best fit. The fitting parameters and $1 \sigma$ and $2 \sigma$ confidence levels (CL) are computed from the Markov chains with 30\% as burn-in. We plotted the histograms and contour plots using the Python package \texttt{GetDist}\cite{lewis2019getdistpythonpackageanalysing}.

\section{Observational Data for the MCMC method}
\label{sec4:obs}

In table~\ref{tab:galaxies}, we present a summary of the characteristics of the galaxies used in the present work, as well as the references from which we obtained the data for this article. In the following subsections, we give more details of the host galaxies studied.

In Appendix~\ref{appendix-priors}, in Table~\ref{tab:priors}, we report the priors used for the MCMC calibration method, for each one of the galaxies and the mSFDM and halo components.

\begin{table*}
    \centering
    \begin{tabular}{c|c|c|c|c|c|c}
    \hline
        Host Galaxy & Morphology & i & $v_\text{max}$ & $D$ & $r_\text{max}$ & Source
         \\
         & & (°) & (km/s) & (Mpc) & (kpc) \\
    \hline
    MW & SB & & 250 & 0.008* & 64.5 & \cite{Sofue:2017} \\
    M31 & Sb* & 72.2* & 254 & 0.75* & 36.5 & \cite{Corbelli:2010} \\
    Cen A & S0* & 45.3* & 257 & 4.0* & 6.3 & \cite{Struve:2010} \\
    NGC 2950 & SB0 & 50 & 156 & 18.9 & 1.01 & \cite{Moiseev:2003qj}\\
    NGC 3245 & S0* & 61.8* & 147 & 20.9* & 1.34 & \cite{Simien:1998} \\
    NGC 3338 & Sc & 55 & 174 & 23.1* & 40.7 & \cite{Rhee:1996} \\
    NGC 3437 & SABc & 69 & 194 & 22.3* & 4.47 & \cite{Courteau:1997} \\
    NGC 5690 & Sc* & 75.9* & 183 & 19.1* & 7.55 & \cite{Courteau:1997} \\
    NGC 5866 & SA0 & 90* & 210 & 14.7* & 3.4 & \cite{Fisher:1997}\\
    NGC 7332 & S0 & 90* & 147 & 23.0* & 5.3 & \cite{Simien:1997} \\
    \hline
    \end{tabular}
    \caption{Characteristics of the galaxies used in the present work. The columns read:
(1)~The name of the galaxy, (2)~Morphology, (3)~Inclination between the line-of-sight and the polar axis of the galaxy, (4)~Maximum velocity of the rotation curve, $v_\text{max}$, (5)~Distance to the galaxy, $D$, (6)~Maximum spatial extent in kpc of either the approaching or receding side of the galaxy, $r_\text{max}$, (7)~Source of the data used in this work (for the data with * the source is HyperLeda database \cite[][\textit{http://leda.univ-lyon1.fr/}]{Makarov:2014} and references therein).}
    \label{tab:galaxies}
\end{table*}

\subsection{Milky Way}
\label{obs:MW}

For the Milky Way, we use the rotation velocities data reported in \cite{Sofue:2017}, obtained for the values at the Sun radius $(R_0,V_0) = (8.0 \text{ kpc}, 238 \text{ km/s})$. Data are reported between 0.29 and 1280 kpc. However, we use the data from 
0.564 to 64.5 kpc, since the errors at larger radii are huge. 
For simplicity, we avoid the innermost region, where the supermassive black hole Sgr A* is dominant, and where it has been found a massive central core (or second bulge) inside 0.1 kpc \citep[see e.g.][]{Maleki:2020}.

The initial values for the MCMC method are taken from those reported in \citep{Sofue:2012}: For the exponential bulge $m_b = (1.652 \pm 0.083) \times 10^{10} M_\odot$ and $h_b = (522 \pm 37)$ pc; for the exponential disc $m_d = (3.41 \pm 0.41) \times 10^{10} M_\odot$ and $h_d = (3.19 \pm 0.35)$ kpc.
The total mass of the Milky Way, including the DM halo up to $\sim 150$ kpc, has been estimated to be $\sim 3 \times 10^{11} M_\odot$ \citep{Sofue:2017}. We did not take into account the gas contribution.

\subsection{Andromeda}
\label{obs:M31}

For Andromeda, we exclude the central region from the analysis, because of the large surface density of baryons and a possible adiabatic contraction there \citep[see][and references therein]{Corbelli:2010}. As the bulge is the most important contribution in that region, we do not fit such a component, we use only an exponential disc, starting from 8.5 kpc. 

We use the data reported in \cite{Corbelli:2010}. Following that work, the disc luminosity $L_d$ used to obtain the mass parameter $m_d$ is $2.1 \times 10^{10} L_\odot$. In their analysis, the authors vary the disc mass-to-light ratios between 2.5 and 8 $M_\odot/L_\odot$. For the best fitting $\Lambda$CDM model, they obtained $(M/L)_d=5.0 M_\odot/L_\odot$, concentration parameter $C=12.0$, and length-scale $h_d=4.5$ kpc (this parameter varies between 4.5 and 6.1 kpc), for an exponential disc. The total bulge + disc mass in that work is $1.3 \times 10^{11} M_\odot$, and the dark NFW halo is $1.2 \times 10^{12} M_\odot$.

According to the last values, we use as initial numbers for the MCMC method: $m_d = (5.25-16.8) \times 10^{10} M_\odot$ and $h_d = 4.5$ kpc. For simplicity, we did not take into account the gas mass contribution.

\subsection{Centaurus A}
\label{obs:CenA}

Centaurus A (NGC 5128) is an active galactic nucleus, with a supermassive black hole at its center ejecting relativistic jets emitting in X-ray and radio wavelengths. It is a peculiar galaxy since it is in collision with another spiral galaxy. For the analysis, we exclude the innermost region and the region where the bulge is dominant, thus we fit an exponential disc with the mSFDM halo.

We use the rotation curve obtained in \citep{Struve:2010}, with ATCA 21-cm line observations of the neutral hydrogen (HI). The HI kinematics shows that the galaxy has a regularly rotating, warped main disc (for $r<6$ kpc). No bar or bulge is necessary to reproduce the HI kinematics. The total extent of the HI disc is $\sim 7.5$ kpc in radius, and its total mass is $4.9 \times 10^8 M_\odot$.

\subsection{MW-like Galaxies with Satellites}
\label{obs:6-galaxies}

In \cite{Nashimoto_2022}, the authors analyzed observations from Subaru/Hyper Suprime-Cam of satellite galaxies around 9 Milky Way-like galaxies located outside of the Local Group. After screening to eliminate false positives, they identified 51 dwarf satellite galaxies within the virial radius of the host galaxies.

The study found that the average luminosity function of the satellite galaxies is consistent with that of the Milky Way satellites, though the luminosity function of each host galaxy varies significantly. They also found that more massive host galaxies tend to have a larger number of satellites. The physical properties of the satellites, such as the size-luminosity relation, were consistent with those of the Milky Way satellites.

However, there was a notable difference in the spatial distribution. The satellite galaxies outside the Local Group showed no signs of concentration or alignment, unlike the MW satellites, which are more concentrated around the host galaxy and exhibit significant alignment.

The study concluded that this difference in spatial distribution might represent a peculiarity of the Milky Way satellites and further research is required to understand the origin of this trend, considering that the observation was focused on relatively massive satellites with $M_V < -10$.

Despite not observing a clear alignment of satellites in those galaxies, we used their reported observations to test the SFDM model with two quantum states and see if there is any indication of a distribution around the atom-like structure of the SF model. While the absence of a pronounced alignment of satellite galaxies might suggest that the model's atom-like structure is not immediately evident, exploring this possible implication of the model is relevant.

Thus, we fit the rotation curves of 6 of the 9 galaxies we found to obtain the best parameters of the model, with a mSFDM halo.

The rotation curve and the data for NGC 2950 are taken from \cite{Moiseev:2003qj}. The line-of-sight velocity data for NGC 3437 are taken from \cite{Courteau:1997}, from which we made the interpolation between the approaching and the receding parts of the galaxy to obtain the rotation curve and errors we used in the present work. The position-velocity map for NGC 3338 is taken from \cite{Rhee:1996}, from which we obtained the rotation curve and errors used in this work. The rotation velocities for NGC 3245 and NGC 7332 are taken from \cite{Simien:1998} and \cite{Simien:1997}, respectively. For NGC 5866 the rotation velocities are taken from \cite{Fisher:1997}, from which we interpolated between the approaching and the receding parts of the galaxy.

\section{Results}
\label{sec5:results}

Remarkably, all the galaxies could be successfully reproduced with the ground and the first excited states of the mSFDM model studied in this work. The best-fit parameters, obtained from the MCMC method implemented in Python (see Section~\ref{sec:mcmc}), are listed in table \ref{tab:BF-results}. The errors are reported to $1\sigma$ of confidence level. We also show the posterior distributions for the corresponding parameters (figures~\ref{fig:BF-MW}-\ref{fig:ngc-corner}).

\subsection{Milky Way}
\label{sub:MW}

The best-fit parameters, obtained for the exponential bulge and disc, and the ground and first excited states of the mSFDM halo, are listed in table \ref{tab:BF-results}.

\begin{figure}
    \centering
    \includegraphics[width=0.49\linewidth]{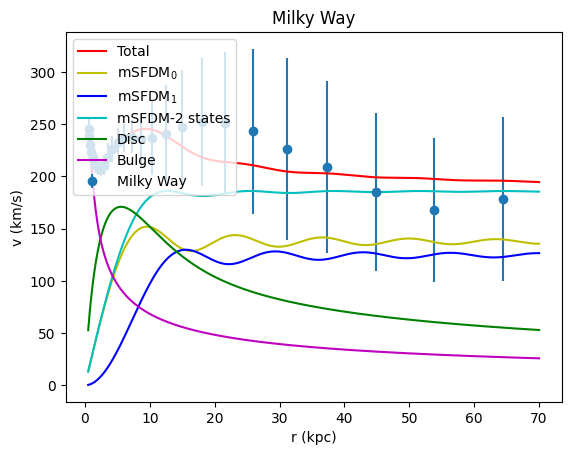}
    \includegraphics[width=0.49\linewidth]{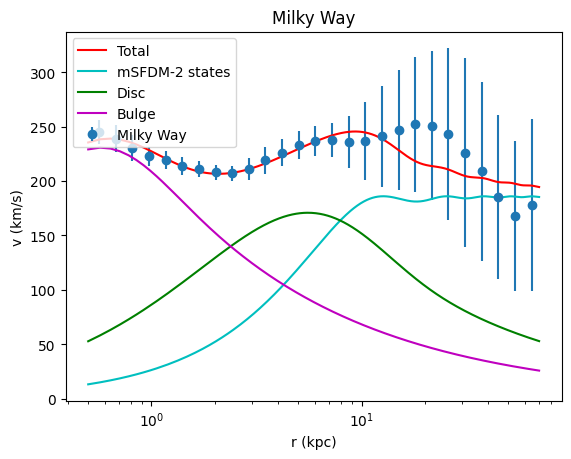}\\
    \includegraphics[width=0.49\linewidth]{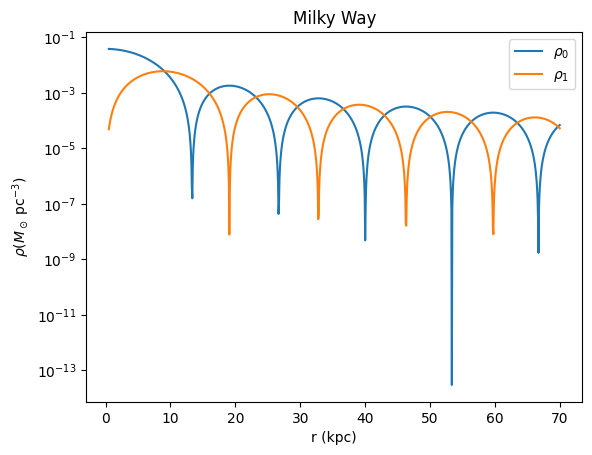}
    \includegraphics[width=0.49\linewidth]{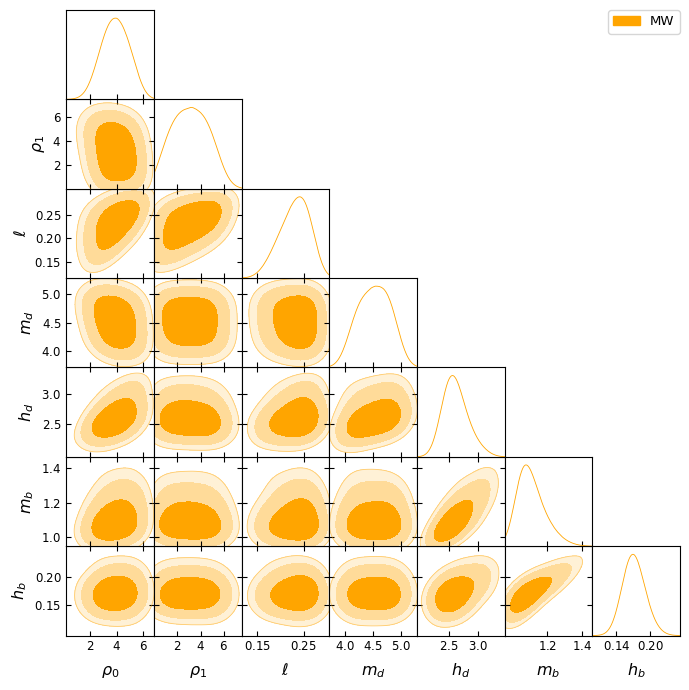}
    \caption{(Top) Best fit results for the Milky Way with bulge, disc, and mSFDM components, in normal and logarithmic scale, the last one showing the central region of the Galaxy. (Bottom) The behavior of the densities, using the same best-fit parameters. Posterior distributions for the parameters $\rho_0(10^{-2}M_\odot/\text{pc}^{-3})$, $\rho_1(10^{-2}M_\odot/\text{pc}^{-3})$, $l(\text{kpc}^{-1})$, $m_b(10^{10}M_\odot)$, $h_b$(kpc), $m_d(10^{10}M_\odot)$, and $h_d$(kpc). The contour lines correspond to the $1\sigma$, $2\sigma$, and $3\sigma$ confidence regions.}
    \label{fig:BF-MW}
\end{figure}

In figure~\ref{fig:BF-MW}, we show the best-fit MW rotation curve up to 64.5 kpc and the densities of the ground and first excited states of the mSFDM configuration. We cut the DM contribution as defined in \eqref{R-max}. The posterior distributions for the 7 free parameters are also shown in this figure.

In table~\ref{tab:BF-mSFDM_R}, we compare the maximum radius of the mSFDM configuration, with the virial radius reported in \citep{Sofue:2012}. Both radii, $R_0^\text{max}$ and $R_1^\text{max}$, are around 0.3 and 0.5 times the virial radius, respectively. With this result, the two-lobe structure, originating in the center of the Milky Way and orthogonal to its plane, would enclose around 53 of 61 satellites listed in \citep{DES:2019vzn}.

\subsection{Andromeda}
\label{sub:M31}

\begin{figure}
    \centering
    \includegraphics[width=0.49\linewidth]{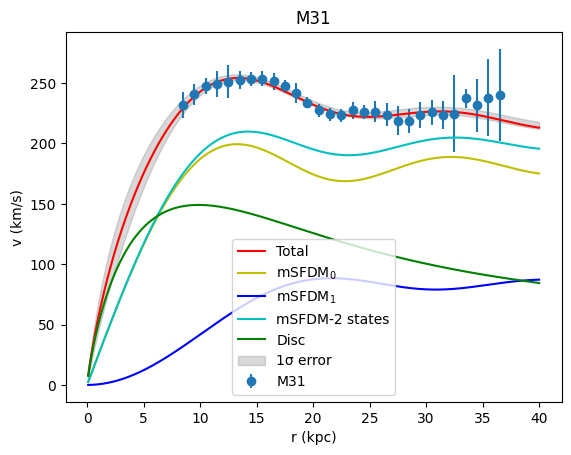}
    \includegraphics[width=0.49\linewidth]{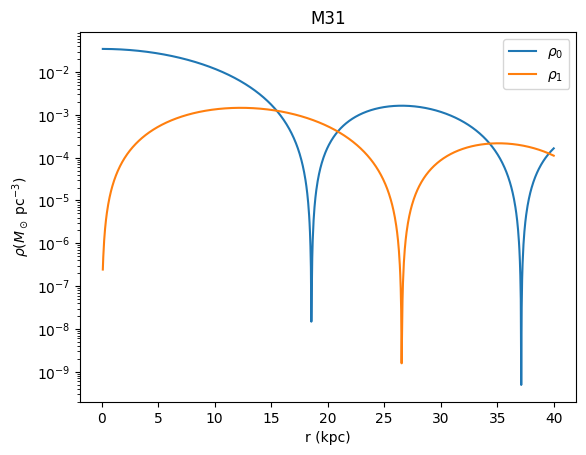}\\
    \includegraphics[width=0.55\linewidth]{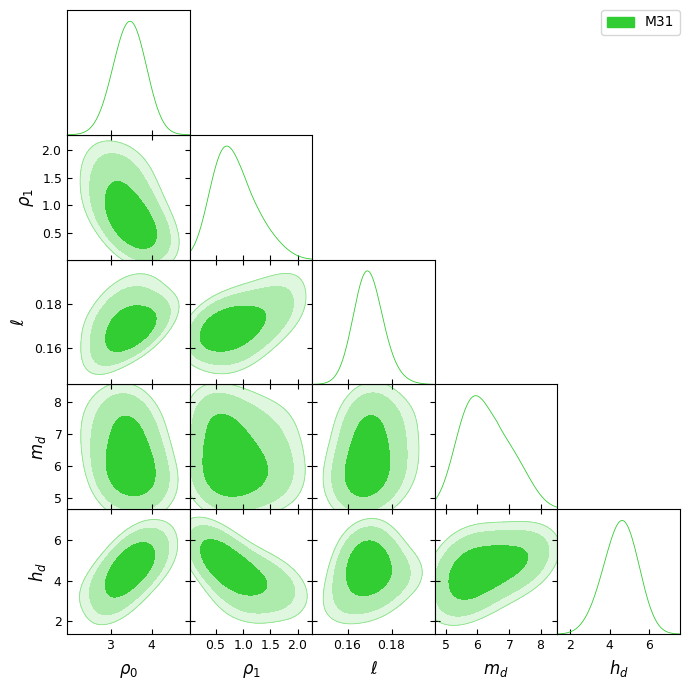}
    \caption{(Top) Best-fit rotation curve of M31 with disc and mSFDM components. Densities of the mSFDM ground and first excited state for the same best-fit parameters. (Bottom) Posterior distributions for the parameters $\rho_0(10^{-2}M_\odot/\text{pc}^{-3})$, $\rho_1(10^{-2}M_\odot/\text{pc}^{-3})$, $l(\text{kpc}^{-1})$, $m_d(10^{10}M_\odot)$, and $h_d$(kpc). The contour lines correspond to the $1\sigma$, $2\sigma$, and $3\sigma$ confidence regions.}
    \label{fig:BF-M31}
\end{figure}

The best-fit parameters for Andromeda for an exponential disc and mSFDM halo are listed in table \ref{tab:BF-results}. In figure~\ref{fig:BF-M31}, we show the best-fit M31 rotation curve up to 36.5 kpc and the densities of the ground and first excited states. We also show the posterior distributions for the 5 free parameters.

In table~\ref{tab:BF-mSFDM_R}, we show the maximum radii of both mSFDM states, the ground and the first excited states, and the virial radius reported in \cite{Corbelli:2010}. Such radii are around 0.75 and 0.5 times the virial radius, respectively. With this result, the two-lobe structure, originating in the center of Andromeda and orthogonal to its plane, would enclose 10 of 18 satellites listed in \cite{Koch:2005kg}.

\subsection{Centaurus A}
\label{sub:CenA}

\begin{figure}
    \centering
    \includegraphics[width=0.49\linewidth]{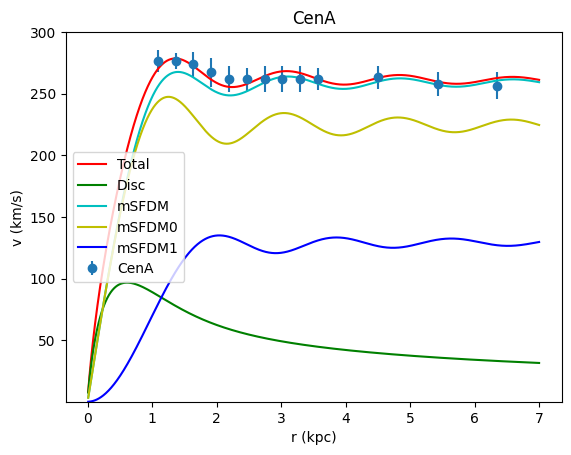}
    \includegraphics[width=0.49\linewidth]{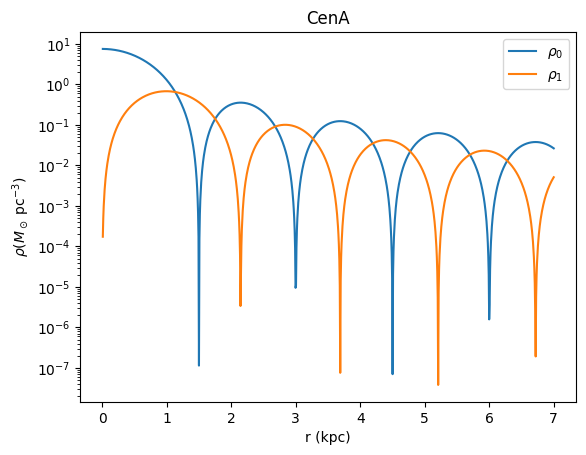}\\
    \includegraphics[width=0.49\linewidth]{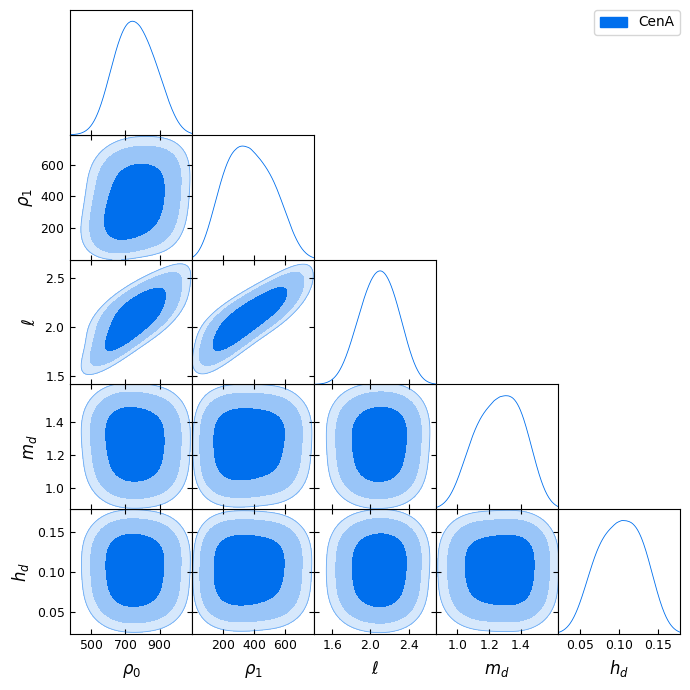}
     \caption{(Top) Best-fit rotation curve of Centaurus A with disc and mSFDM components. Densities of the mSFDM ground and first excited state for the same best-fit parameters. (Bottom) Posterior distributions for the parameters $\rho_0(10^{-2}M_\odot/\text{pc}^{-3})$, $\rho_1(10^{-2}M_\odot/\text{pc}^{-3})$, $l(\text{kpc}^{-1})$, $m_d(10^{10}M_\odot)$, and $h_d$(kpc). The contour lines correspond to the $1\sigma$, $2\sigma$, and $3\sigma$ confidence regions.}
    \label{fig:BF-CenA}
\end{figure}

\begin{figure}
    \centering
    \includegraphics[width=0.46\linewidth]{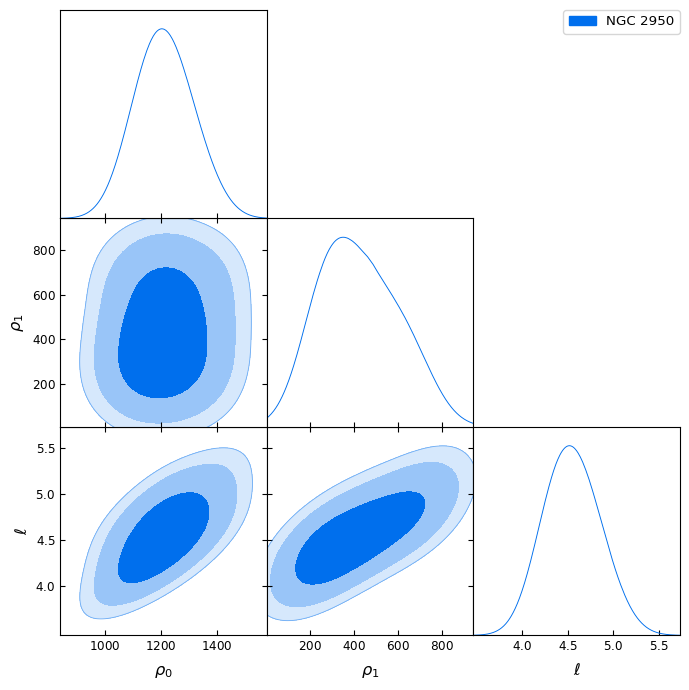}
    \includegraphics[width=0.46\linewidth]{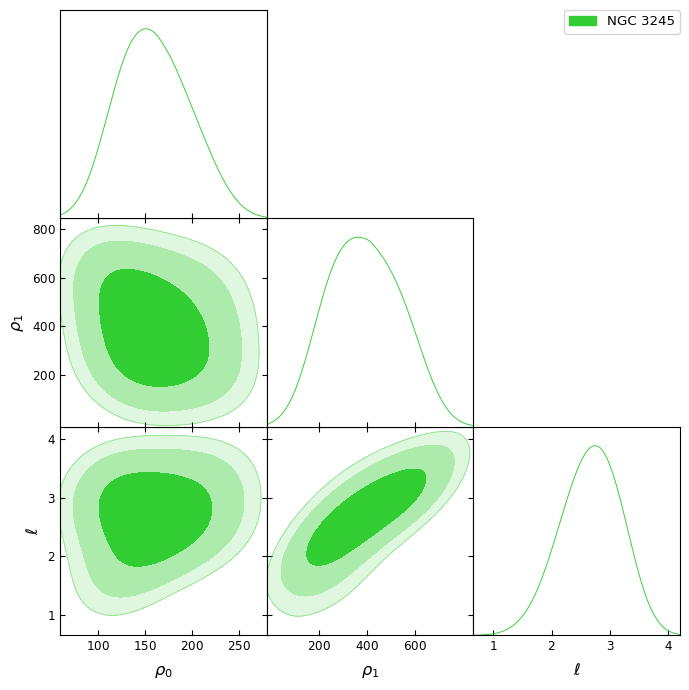}\\
        \includegraphics[width=0.46\linewidth]{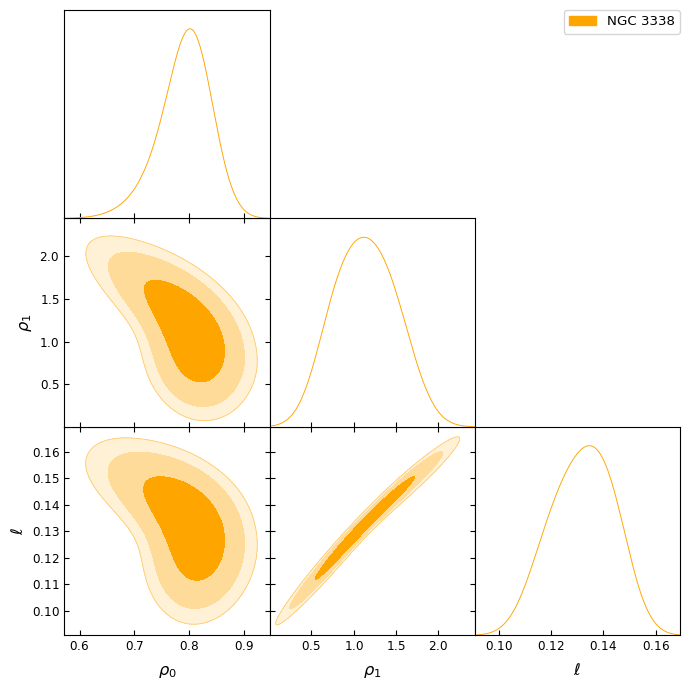}
    \includegraphics[width=0.46\linewidth]{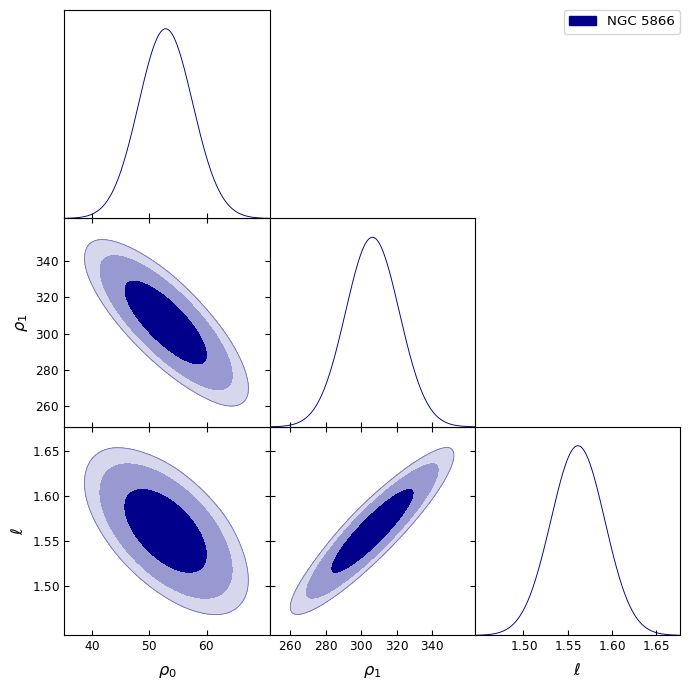}\\
    \includegraphics[width=0.46\linewidth]{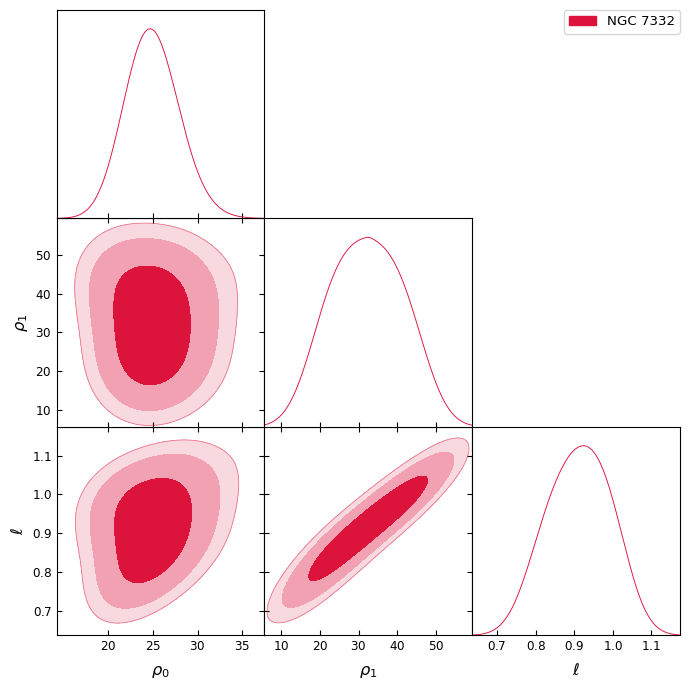}
    \includegraphics[width=0.5\linewidth]{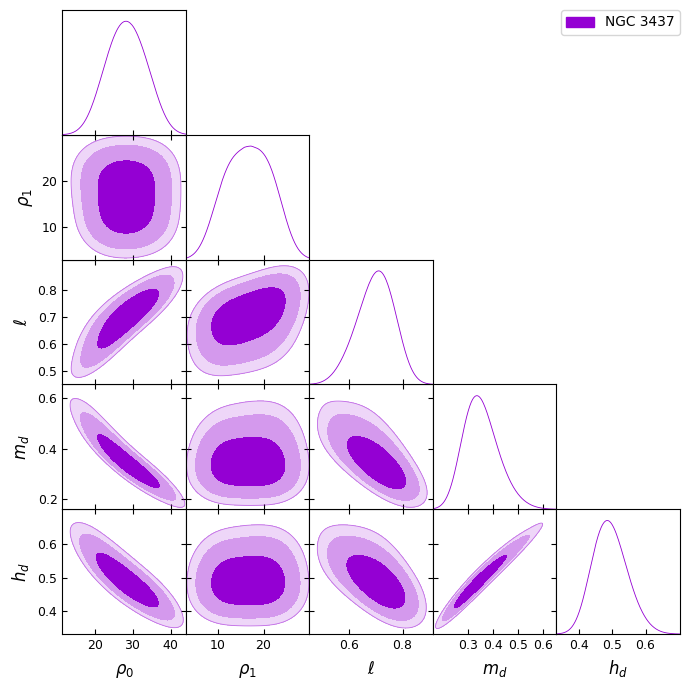}
    \caption{Posterior distributions for the parameters $\rho_0(10^{-2}M_\odot/\text{pc}^{-3})$, $\rho_1(10^{-2}M_\odot/\text{pc}^{-3})$, and $l(\text{kpc}^{-1})$, for the galaxies with mSFDM-only, NGC 2950, 3245, 3338, 5866 and 7332. In the case of NGC 3437, the contours include $m_d(10^{10}M_\odot)$ and $h_d$(kpc), the parameters of the exponential disc.  The contour lines correspond to the $1\sigma$, $2\sigma$, and $3\sigma$ confidence regions.}
    \label{fig:ngc-corner}
\end{figure}

The best-fit parameters, obtained from the MCMC method for an exponential disc and mSDFM halo, are listed in table~\ref{tab:BF-results}. Figure~\ref{fig:BF-CenA} shows the resulting best-fit rotation curve and the behavior of the SF densities. Notice that the best-fit rotation curve agrees well with the observations. We also show the resulting posterior distributions for the 5 free parameters.

In table~\ref{tab:BF-mSFDM_R}, we show the maximum radii of both mSFDM states, the ground and the first excited states, and the virial radius reported in \cite{Muller:2019}. Such radii are around 0.7 and 0.4 times the virial radius, respectively. With this result, the two-lobe structure, originating in the center of Centaurus A and orthogonal to its plane, would enclose 15 of 31 satellites (5 with velocity measurements and 10 without velocities), and mainly the whole structure of 1239 planetary nebulae reported in \cite{Muller:2018hks} (see the central squared region of figure 1 in that work).

\subsection{MW-like Galaxies with Satellites}
\label{6-galaxies}

\begin{figure*}
    \centering
    \includegraphics[width=0.49\textwidth]{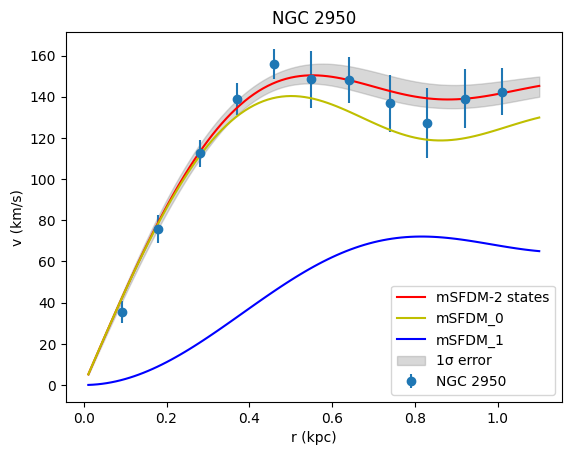}
    \includegraphics[width=0.49\textwidth]{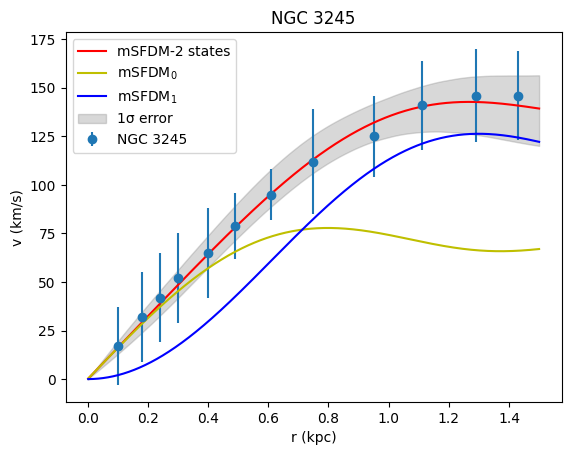}
    \caption{Rotation curves of galaxies NGC 2950 and NGC 3245 using the ground state and the first excited state of the SFDM model. Note that the fit of the observed rotation curves agrees well with the model.}
\label{fig:2950-3245}
\end{figure*}

\begin{figure*}
    \centering
    \includegraphics[scale=0.6]{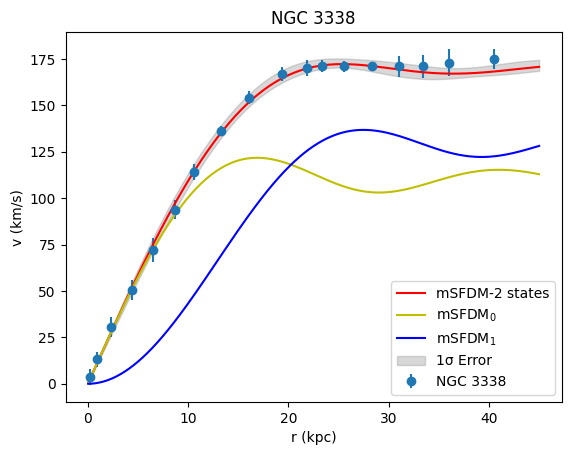}
    \includegraphics[scale=0.26]{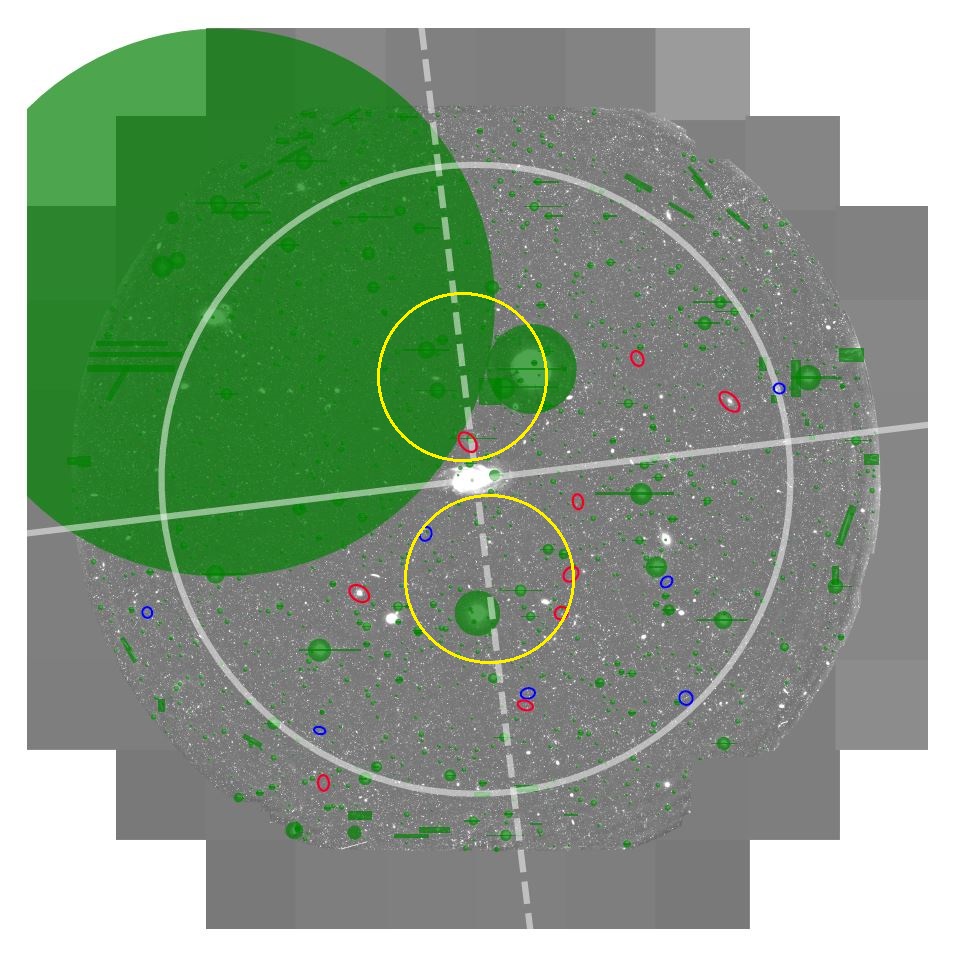}\\
    \includegraphics[scale=0.6]{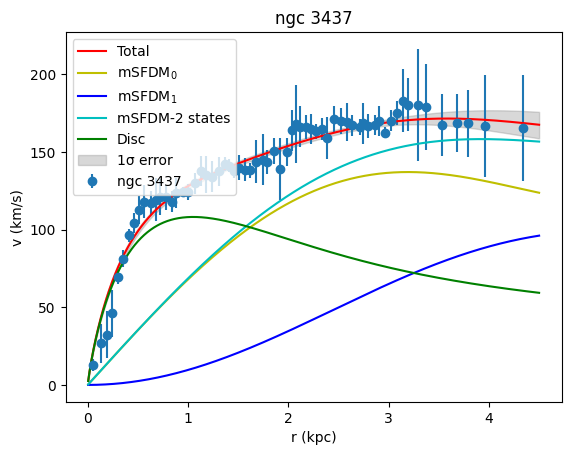}
    \includegraphics[scale=0.26]{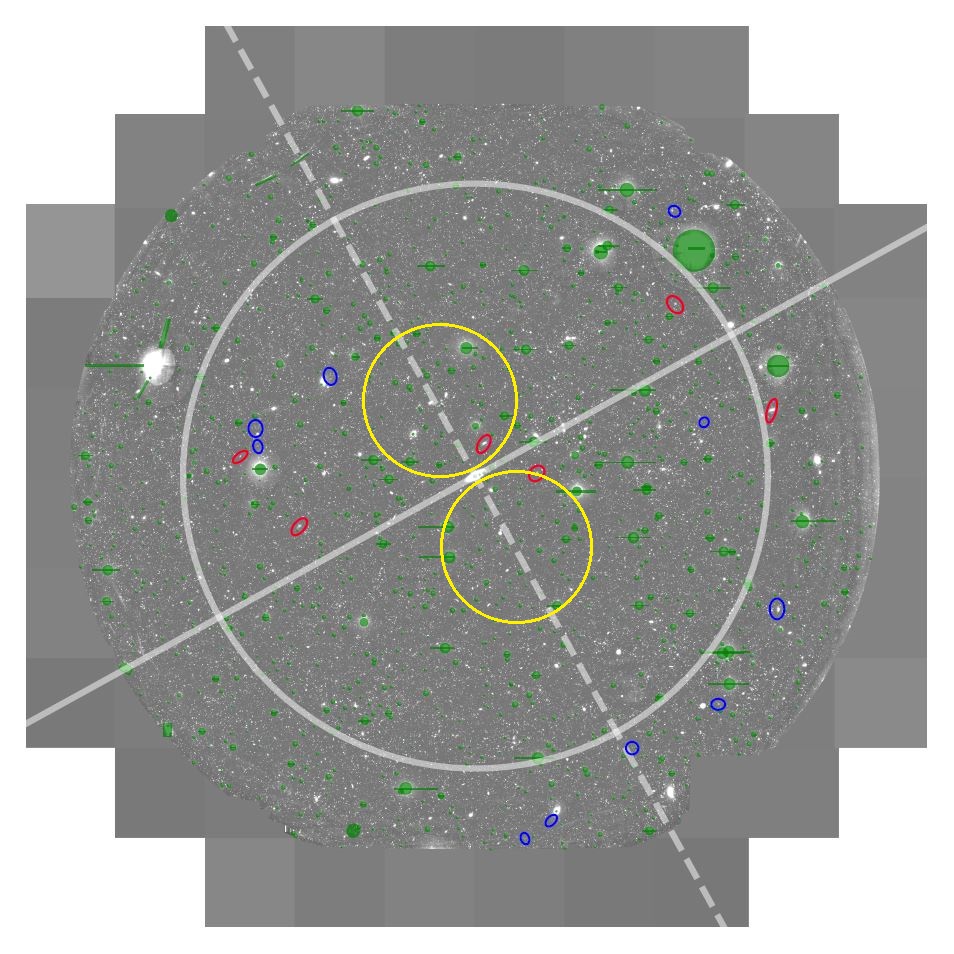}
    \caption{(Left) Rotation curves of galaxies NGC 3388 and NGC 3437 using the ground and first excited states of the mSFDM model. The fit of the observed rotation curves agrees well with the model. (Right) The same galaxies showing their satellite galaxies reproduced from \citep{Nashimoto_2022}, with permission of the authors. The red and blue circles in the original figures represent the secure and possible dwarf galaxies, respectively. The green shaded circles and boxes show the masked regions. The big white circle, also in the original figures, is the virial radius of the host galaxy.
    In each galaxy, we superposed the yellow solid circles, which show the two-lobe excited state of the mSFDM.}
    \label{fig:3338-3437}
\end{figure*}

\begin{figure*}
    \centering
    \includegraphics[scale=0.6]{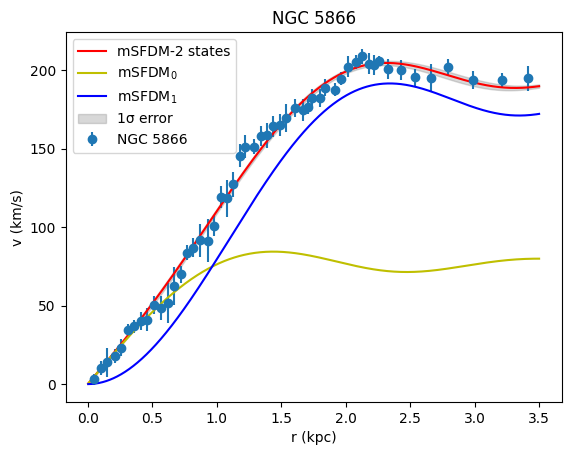}
    \includegraphics[scale=0.26]{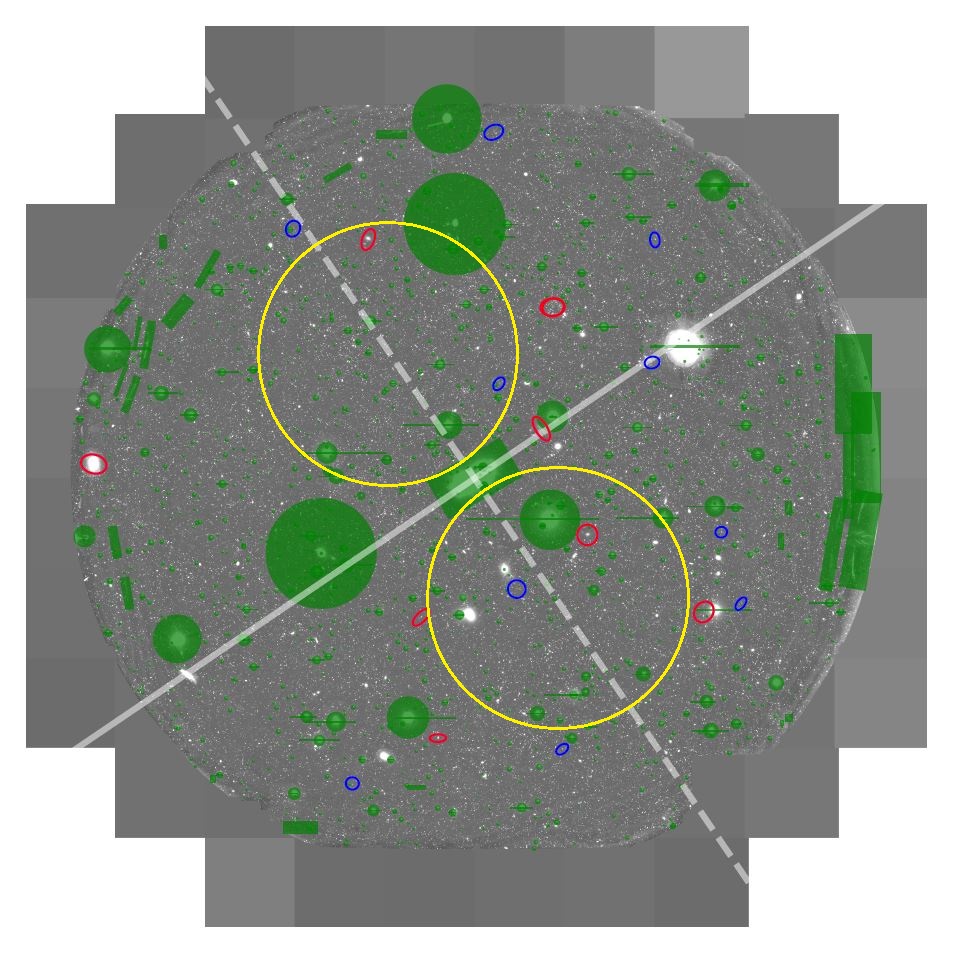}\\
    \includegraphics[scale=0.6]{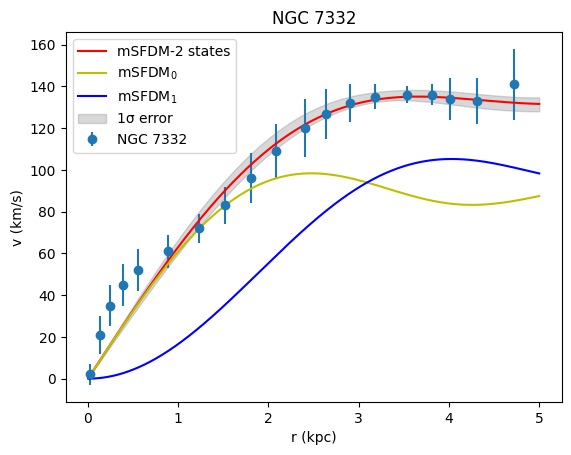}
    \includegraphics[scale=0.26]{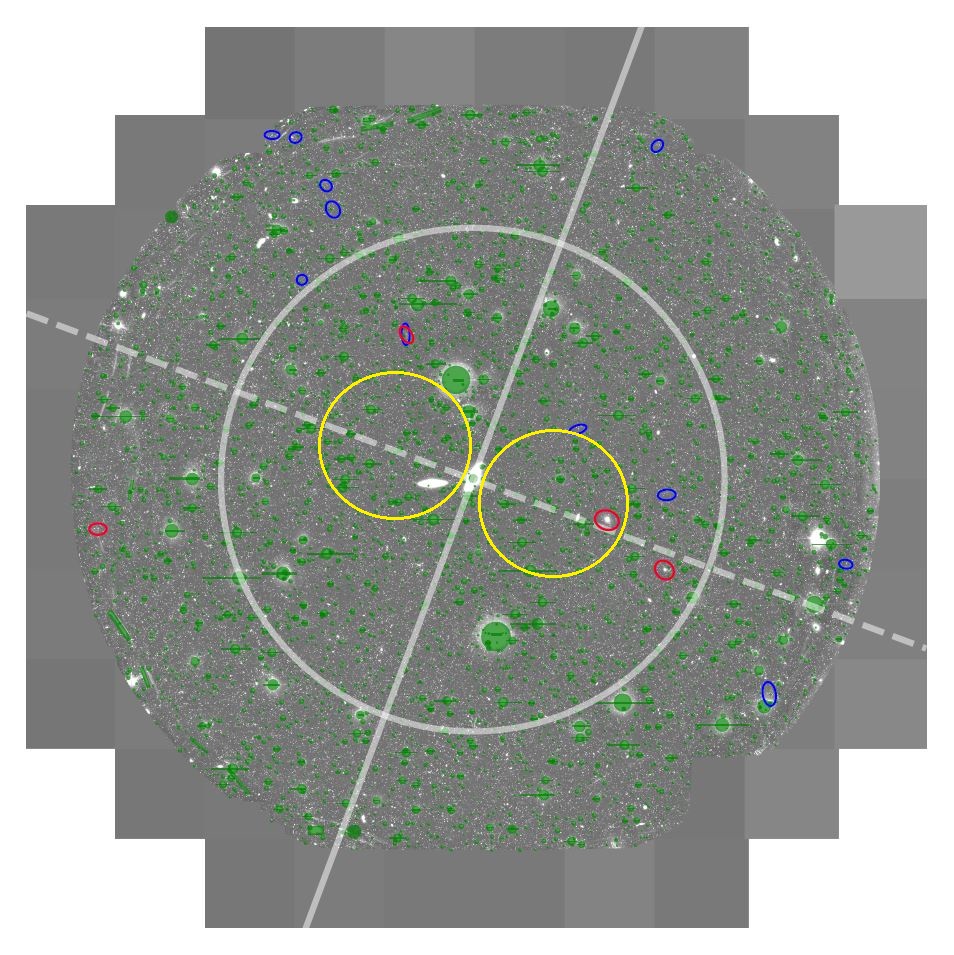}
    \caption{(Left) Rotation curves of galaxies NGC 3388 and NGC 3437 using the ground and first excited states of the mSFDM model. The fit of the observed rotation curves agrees well with the model. (Right) The same galaxies showing their satellite galaxies reproduced from \citep{Nashimoto_2022}, with permission of the authors. The red and blue circles in the original figures represent the secure and possible dwarf galaxies, respectively. The green shaded circles and boxes show the masked regions. The big white circle, also in the original figures, is the virial radius of the host galaxy; for NGC 5866 is outside the observational region.
    In each galaxy, we superposed the yellow solid circles, which show the two-lobe excited state of the mSFDM.}
    \label{fig:5866-7332}
\end{figure*}

For these galaxies, we used a mSFDM halo and only for NGC 3437 we included an exponential disc to better fit the data. The results of these fits are presented in table~\ref{tab:BF-results}. Figures~\ref{fig:2950-3245}-\ref{fig:5866-7332} show the best-fit rotation curves for each galaxy, and the figures reproduced from \citep{Nashimoto_2022} for NGC 3338, 3437, 5866 and 7332, show the central galaxy and the confirmed and tentative satellites within the virial radius of each galaxy. We superposed with yellow solid circles the resulting two-lobe structure (the excited state) from our best fits for the host galaxies, with radius $R_\text{max}^1$ (see table~\ref{tab:BF-mSFDM_R}).

Although the alignment in all the galaxies in \citep{Nashimoto_2022} is not clear, in some cases it would be possible to explain that dwarf galaxies are found within the gravitational potential traced by the two-lobe structure of the first excited state, as in NGC 3338 (the galaxies near the center), two dwarf galaxies in NGC 3437 near the center, about 7 satellite galaxies in NGC 5866, and one or two galaxies around NGC 7332. The corresponding figures for NGC 2950 and NGC 3245 can be found in \cite{Tanaka:2018}; in these cases, for NGC 2950 it is possible to find 3 galaxies inside the gravitational potential of the lobes, and in NGC 3245 about 6 galaxies.

\begin{table*}
    \centering
    \begin{tabular}{cccccc}
    \hline
        \multicolumn{6}{c}{mSFDM halo results}\\
    \hline
        Host Galaxy & $\rho_0$ & $\rho_1$ & $l$ & $M_\text{halo}$ & $-2 \ln \mathcal{L}$ \\
         & ($10^{-2} M_\odot \text{pc}^{-3}$) & ($10^{-2} M_\odot \text{pc}^{-3}$) & ($\text{kpc}^{-1}$) & $(10^{12} M_\odot)$ & \\
    \hline
    MW & $3.88^{+1.16}_{-1.08}$ & $3.19^{+1.79}_{-1.69}$ & $0.23^{+0.03}_{-0.03}$ & 1.38 & 2.7\\
    M 31 & $3.46^{+0.38}_{-0.40}$ & $0.766^{+0.477}_{-0.323}$ & $0.169^{+0.007}_{-0.006}$ & 1.8 & 7.22 \\
    Cen A & $750.77^{+133.42}_{-115.18}$ & $356.31^{+193.57}_{-153.68}$ & $2.09^{+0.20}_{-0.20}$ & 3.7 & 3.98\\
    NGC 2950 & $1204.8^{+108.74}_{-101.07}$ &	$395.62^{+240.39}_{-171.03}$ & $4.52^{+0.32}_{-0.30}$ & 0.674 &  4.78\\
    NGC 3245 & $155.63^{+44.5}_{-38.0}$ & $383.03^{+179.89}_{-163.9}$ & $2.70^{+0.50}_{-0.56}$ & 0.598 & 0.16\\
    NGC 3338 & $0.79^{+0.03}_{-0.04}$ & $1.13^{+0.41}_{-0.41}$ & $0.13^{+0.01}_{-0.01}$ & 1.06 & 5.11\\
    NGC 3437 & $28.26^{+5.56}_{-5.42}$ & $16.78^{+5.67}_{-5.93}$ & $0.70^{+0.06}_{-0.07}$ & 0.789 & 38.33\\
    NGC 5866 & $52.87^{+4.54}_{-4.58}$ & $306.2^{+14.62}_{-14.72}$ & $1.56^{+0.02}_{-0.02}$ & 1.84 & 39.02\\
    NGC 7332 & $24.69^{+3.05}_{-2.78}$ & $32.05^{+11.05}_{-10.96}$ & $0.91^{+0.08}_{-0.09}$ & 0.502 & 13.48 \\
    \hline
    \end{tabular}
    \begin{tabular}{ccccc}
        \multicolumn{5}{c}{Baryonic components results}\\
    \hline
    & \multicolumn{2}{c}{Disc} & \multicolumn{2}{c}{Bulge} \\
    Host Galaxy & $m_d$ & $h_d$ & $m_b$ & $h_b$ \\
    & ($10^{10} M_\odot$) & (kpc) & ($10^{10} M_\odot$) & (pc) \\
    \hline
    MW & $4.53^{+0.32}_{-0.34}$ & $2.58^{+0.23}_{-0.17}$ & $1.08^{+0.09}_{-0.06}$ & $0.17^{+0.02}_{-0.01}$ \\
    M31 & $6.13^{+1.02}_{-0.67}$ & $4.59^{+0.81}_{-0.91}$ & -- & -- \\
    Cen A & $0.012^{+0.001}_{-0.001}$ & $0.10^{+0.032}_{-0.035}$ & -- & -- \\
    NGC 3437 & $0.34^{+0.07}_{-0.06}$ & $0.48^{+0.05}_{-0.04}$ & -- & -- \\
    \hline
    \end{tabular}
    \caption{Best-fit results from the MCMC method. (Top) Results for the mSFDM halos reported with $1\sigma$ errors. The halo mass, $M_\text{halo}$, is calculated from the sum of the ground and excited states, up to $R^\text{max}$, reported in table~\ref{tab:BF-mSFDM_R}. We report the minimum values $-2\ln \mathcal{L}$ from the MCMC method. (Bottom) Results of the baryonic components reported with $1\sigma$ errors. To fit the MW rotation curve, we use an exponential bulge and disc. For M31, Cen A, and NGC 3437, we use an exponential disc.}
    \label{tab:BF-results}
\end{table*}

\begin{table}
    \centering
    \begin{tabular}{ccccc}
    \hline
    \multicolumn{5}{c}{Radii for the mSFDM configurations}\\
    \hline
        Host Galaxy & $R_0^\text{max}$ & $R_1^\text{max}$ & $R_\text{vir}$ & $R_1^\text{max}/R_\text{vir}$\\
         & (kpc) & (kpc) & (kpc) & \\
    \hline
    MW & 170.0 & 159.0 & 385.0 & 0.41\\
    M 31  & 213.0 & 111.0 & 270.0 & 0.41\\
    Cen A & 257.5 & 180.0 & 409.0 &  0.44\\
    NGC 2950 & 153.7 & 86.5 & 176.0 &  0.49\\
    NGC 3245 & 92.2 & 143.9 & 226.7 &  0.63\\
    NGC 3338 & 130.0 & 166.0 & 266.4 &  0.62\\
    NGC 3437 & 149.7 & 116.0 & 243.0 &  0.47\\
    NGC 5866 & 91.0 & 217.5 & 301.2 &  0.72\\
    NGC 7332 & 108.0 & 124.0 & 210.0 &  0.59\\
    \hline
    \end{tabular}
    \caption{Radii for the mSFDM halos. The maximum radii of both mSFDM states are obtained such that $\rho_i(R_i^\text{max})=200\rho_c$ (cf.~equation~\eqref{R-max}), for the densities given by equations~\eqref{eq:rho0} and \eqref{eq:rho1}. The virial radius for the MW is taken from \citep{Sofue:2012}, for M31 is from \citep{Corbelli:2010}, and for Cen A is taken from \citep{Muller:2019}. The virial radius for the last 6 galaxies is taken from \citep{Nashimoto_2022}. We show also the ratio between $R_1^\text{max}$, the radius of the two-lobe structure, and $R_\text{vir}$.}
    \label{tab:BF-mSFDM_R}
\end{table}

\vspace{1cm}

The whole analysis in this Section confirms that the mSFDM model with two states is capable of accurately describing the rotation curves of the galaxies under study. The fact that all galaxies were well-fitted with the same number of parameters further supports the robustness and universality of the model in capturing the essential features of the observed rotation curves.

This is a significant advantage over the finite temperature model with real SF presented in \citep{Robles:2012kt}, which required various combinations of ground plus excited states to reproduce different observations \citep{Robles:2012kt,Bernal:2016lll,Bernal:2017oih}. In the present model, with a complex SF and by using the spherical ground state and the two-lobe first excited state only, it is possible to reproduce all the rotation curves of the studied galaxies. In other words, with the spherical structure of the ground state and the two-lobe structure of the first excited state, it is sufficient to reproduce the observations and also explain the gravitational potential responsible for aligning the satellite galaxies as observed in MW, M31, and CenA, and possibly the galaxies in \citep{Nashimoto_2022}.

\section{Conclusions}
\label{sec:conclusions}

The SFDM model has been very successful in explaining several characteristics of DM, which other models cannot explain or need extra physics. The field equations of the SFDM are the system of Schr\"odinger-Poisson equations. In this work, we consider the quantum character of this model to explain the non-homogeneous distribution of satellite galaxies around their host galaxies. This phenomenon has been observed in the Milky Way, Andromeda, and Centaurus A, where the behavior of the satellite galaxies has been well studied. We started from a complex SF in thermal equilibrium with other components of the SM, at the very early stages of the Universe. Then we discuss the conditions for an SSB to take place as the Universe cools down. Then we solve the evolution equations in an FLRW Universe, separating the SF function into its background and linear perturbations, and show the Newtonian limit to find the analytic solutions for the density. Then we solve the Schr\"odinger equation by using separation of variables, where the angular dependence is given by the spherical harmonic functions, which contain axial symmetric terms. We use such terms to model the inhomogeneous behavior of the DM and fit the free parameters of the solutions using the rotation curves of the Milky Way, Andromeda, Centaurus A, and 6 other MW-like galaxies as the host galaxies of dwarf satellites. With this, we can explain the anisotropic distribution of satellite galaxies around their host galaxy, with the two-lobe shape of the first excited state of the multistate SFDM, from complex SF within the finite temperature scenario.

This is the first natural explanation for the anomalous behavior of the satellite galaxies, with the surprising result that the first three galaxies where satellite galaxies can be seen with good resolution, have the same behavior. If in the future we can see more galaxies where their satellite galaxies maintain the same behavior, this will be a strong support for the multistate SFDM paradigm.

We also find that the length-scale $l$ of the host galaxy depends on the energy of the vibrations $\omega$ and the effective mass $m_\Phi$. The SFDM collapses after spinning and this collapse causes the temperature of the SFDM to rise again. The final temperature therefore determines the effective mass of the SF, and this mass determines the scale of the galaxy. This is why each galaxy has different $l$ scales. As we have seen, this scale is determined by $l^2=\omega^2-m_\Phi^2 a^2$. If we take the self-interaction constant $\lambda$ as in \cite{Li:2013nal}, the critical temperature $T_c$ is very high and the effective mass $m_\Phi=m$, is the mass of the scalar field. Therefore, the parameter $\omega$ is simply $\omega^2=l^2+m^2 a^2\sim m^2$, since $l\sim 10^{-23}$eV.
Using the values found with the rotation curves, we can obtain the size of the excited states that can explain the VPOS in these galaxies. The fact that the scale $l$ depends on the galaxy we are studying explains why some observations fit the SFDM mass to one value and different observations of galaxies give a different value, since the scale is not determined by the SF mass, but by the scale $l$ that depends on the final collapse temperature of each galaxy.

\section*{Acknowledgements}

We gratefully acknowledge M.~Nashimoto for permitting us to reproduce the figures included in Figs.~\ref{fig:3338-3437} and \ref{fig:5866-7332}. We gratefully appreciate helpful comments and discussion of this work with M.S.~Pawlowski, and of the anonymous referee who revised the previous version carefully and helped us to strengthen the results in the last version of this work.

We acknowledge the usage of the HyperLeda database (http://leda.univ-lyon1.fr).

This work was partially supported by CONAHCyT M\'exico under grants  A1-S-8742, Ciencia de Frontera No.~304001, 376127, 240512. Also by the FORDECYT-PRONACES grant No.~490769, and by the grant I0101/131/07 C-234/07 of the Instituto Avanzado de Cosmolog\'ia (IAC) collaboration (http://www.iac.edu.mx/), and Xiuhcoatl and Abacus clusters at Cinvestav, IPN.


\bibliographystyle{JHEP}
\bibliography{main.bib}

\appendix
\section{Priors for the MCMC method}
\label{appendix-priors}

In table~\ref{tab:priors}, we report the priors used in the MCMC calibration method.

\begin{table*}
    \centering
    \begin{tabular}{cccc}
    \hline
    \hline
        \multicolumn{4}{c}{mSFDM halo priors}\\
    \hline
        Host Galaxy & $\rho_0$ & $\rho_1$ & $l$ \\
          & ($10^{-2} M_\odot \text{pc}^{-3}$) & ($10^{-2} M_\odot \text{pc}^{-3}$) & ($\text{kpc}^{-1}$) \\
    \hline
    MW & $\mathcal{U}(0.5, 6.0)$ & $\mathcal{U}(0.5, 6.0)$ & $\mathcal{U}(0.14, 0.28)$ \\
    M 31 & $\mathcal{U}(2.0, 5.0)$ & $\mathcal{U}(0.3, 2.0)$ & $\mathcal{U}(0.1, 0.2)$ \\
    Cen A & $\mathcal{U}(100.0, 1000)$ & $\mathcal{U}(100.0, 650.0)$ & $\mathcal{U}(0.1, 10.0)$ \\
    NGC 2950 & $\mathcal{U}(150.0, 1500)$ &	$\mathcal{U}(150.0, 800.0)$ & $\mathcal{U}(0.01, 8.0)$\\
    NGC 3245 & $\mathcal{U}(90.0, 250.0)$ & $\mathcal{U}(130.0, 700.0)$ & $\mathcal{U}(1.0, 3.8)$ \\
    NGC 3338 & $\mathcal{U}(0.01, 5.0)$ & $\mathcal{U}(0.01, 5.0)$ & $\mathcal{U}(0.01, 2.0)$\\
    NGC 3437 & $\mathcal{U}(15.0, 40.0)$ & $\mathcal{U}(8.0, 25.0)$ & $\mathcal{U}(0.5, 2.2)$ \\
    NGC 5866 & $\mathcal{U}(0.01, 100.0)$ & $\mathcal{U}(100.0, 500.0)$ & $\mathcal{U}(1.0, 5.0)$\\
    NGC 7332 & $\mathcal{U}(15.0, 50.0)$ & $\mathcal{U}(15.0, 50.0)$ & $\mathcal{U}(0.1, 3.0)$\\
    \end{tabular}
    \begin{tabular}{ccccc}
    \hline
    \hline
        \multicolumn{5}{c}{Baryonic components priors}\\
    \hline
    & \multicolumn{2}{c}{Disc} & \multicolumn{2}{c}{Bulge} \\
    Host Galaxy & $m_d$ & $h_d$ & $m_b$ & $h_b$ \\
    & ($10^{10} M_\odot$) & (kpc) & ($10^{10} M_\odot$) & (pc) \\
    \hline
    MW & $\mathcal{U}(4.0, 5.0)$ & $\mathcal{U}(0.01, 3.3)$ & $\mathcal{U}(1.0, 2.0)$ & $\mathcal{U}(0.1, 0.3)$ \\
    M31 & $\mathcal{U}(5.2, 8.0)$ & $\mathcal{U}(2.0, 7.0)$ & -- & -- \\
    Cen A &$\mathcal{U}(0.01, 0.015)$ & $\mathcal{U}(0.05, 0.15)$ & -- & -- \\
    NGC 3437 & $\mathcal{U}(0.1, 0.7)$ & $\mathcal{U}(0.3, 0.7)$ & -- & -- \\
    \hline
    \end{tabular}
    \caption{\textbf{Priors used in the MCMC code for each galaxy.}}
    \label{tab:priors}
\end{table*}

\end{document}